\documentclass{aastex61}

\usepackage{amsmath}

\received{January 6, 2017}
\revised{}
\accepted{}
\submitjournal{ApJ}
\shorttitle{Observation of the Crab Nebula with HAWC}
\shortauthors{Albert et al.}

\begin{document}

\title{Observation of the Crab Nebula with the HAWC Gamma-Ray Observatory}

\correspondingauthor{John Pretz}
\email{john.pretz@gmail.com}

\author{A.U.~Abeysekara}
\affiliation{Department of Physics and Astronomy, University of Utah, Salt Lake City, UT, USA }

\author{A.~Albert}
\affiliation{Physics Division, Los Alamos National Laboratory, Los Alamos, NM, USA }

\author{R.~Alfaro}
\affiliation{Instituto de F\'{i}sica, Universidad Nacional Aut\'{o}noma de M\'{e}xico, Ciudad de M\'{e}xico, M\'{e}xico }

\author{C.~Alvarez}
\affiliation{Universidad Aut\'{o}noma de Chiapas, Tuxtla Guti\'{e}rrez, Chiapas, M\'{e}xico}

\author{J.D.~\'{A}lvarez}
\affiliation{Universidad Michoacana de San Nicol\'{a}s de Hidalgo, Morelia, M\'{e}xico }

\author{R.~Arceo}
\affiliation{Universidad Aut\'{o}noma de Chiapas, Tuxtla Guti\'{e}rrez, Chiapas, M\'{e}xico}

\author{J.C.~Arteaga-Vel\'{a}zquez}
\affiliation{Universidad Michoacana de San Nicol\'{a}s de Hidalgo, Morelia, M\'{e}xico }

\author{H.A.~Ayala Solares}
\affiliation{Department of Physics, Michigan Technological University, Houghton, MI, USA }

\author{A.S.~Barber}
\affiliation{Department of Physics and Astronomy, University of Utah, Salt Lake City, UT, USA }

\author{N.~Bautista-Elivar}
\affiliation{Universidad Politecnica de Pachuca, Pachuca, Hgo, M\'{e}xico }

% ...

\begin{abstract}

The Crab Nebula is the brightest TeV gamma-ray source in the sky and 
has been used for the past 25 years as a reference source in TeV
astronomy, for calibration and verification of new TeV instruments. 
The High Altitude Water Cherenkov Observatory (HAWC), 
completed in early 2015, has been used to
observe the Crab Nebula at high significance
across nearly the full spectrum of energies
to which HAWC is sensitive. 
HAWC is
unique 
for its wide field-of-view, nearly 2 sr at any instant, and its 
high-energy reach, up to 100 TeV. 
HAWC's sensitivity improves with the gamma-ray
energy. Above $\sim$1 TeV the sensitivity is driven by the
best background rejection and angular resolution 
ever achieved for a wide-field ground array.

We present a time-integrated 
analysis of the Crab using 507 live days of HAWC data
from 2014 November to 2016 June. The spectrum of the 
Crab is fit to a function of the form
$\phi(E) = \phi_0 (E/E_{0})^{-\alpha -\beta\cdot{\rm{ln}}(E/E_{0})}$.
The data is well-fit with values 
of
$\alpha=2.63\pm0.03$, 
$\beta=0.15\pm0.03$, and
log$_{10}(\phi_0~{\rm{cm}^2}~{\rm{s}}~{\rm{TeV}})=-12.60\pm0.02$
when 
$E_{0}$ is fixed at 7 TeV and the fit applies
between 1 and 37 TeV. 
Study of the systematic 
errors in this HAWC measurement is discussed and estimated to 
be $\pm$50\% in the photon flux between 1 and 37 TeV. 

Confirmation of the Crab flux serves to establish the HAWC instrument's 
sensitivity for surveys of the sky. 
The HAWC 
survey will exceed sensitivity of current-generation observatories
and open a new view of 2/3 of the sky above 10 TeV. 

\end{abstract}


\section{Introduction}
\label{sec:intro}

The Crab Pulsar Wind Nebula (the Crab Nebula or the Crab)
occupies a place of special distinction in the 
history of high-energy astrophysics. It was the first high-confidence
TeV detection in 1989 using the Whipple telescope \citep{whipplecrabdiscovery}
and is the brightest steady source in the Northern sky above 1 TeV.  
It has been observed with imaging atmospheric Cherenkov
telescopes (IACTs) since 
\citep{cangaroocrab,hegracrab,hesscrab,veritascrab,magiccrabnebula}. 
The first
observation using a ground array was the 2003 Milagro detection \citep{milagrocrab},
and the signal was subsequently seen in other ground arrays
\citep{tibetcrab,argocrab}.

The TeV emission arises from inverse-Compton (IC) up-scattering of 
low-energy
ambient photons by energetic electrons 
accelerated in shocks surrounding the central pulsar \citep{crabinversecompton}.
Photons from synchrotron emission of the electrons themselves are likely
the dominant IC target with sub-dominant contributions from the 
cosmic microwave background and the extragalactic background light 
\citep{crabmodeling}. 
Despite
rare flaring emission below 1 TeV
\citep{agilecrabflare,fermiflare}, and a potential 
TeV flare \citep{argoflare},
the Crab is generally believed to be steady at
higher energies \citep{hesscrabflarelimits,veritascrabflarelimits,argocrab}.
Consequently, the Crab Nebula has been adopted
as the reference source in TeV astronomy and is a reliable beam of 
high-energy photons to use for calibrating and understanding new TeV 
gamma-ray instruments.

\begin{figure}
\includegraphics[width=0.48\textwidth]{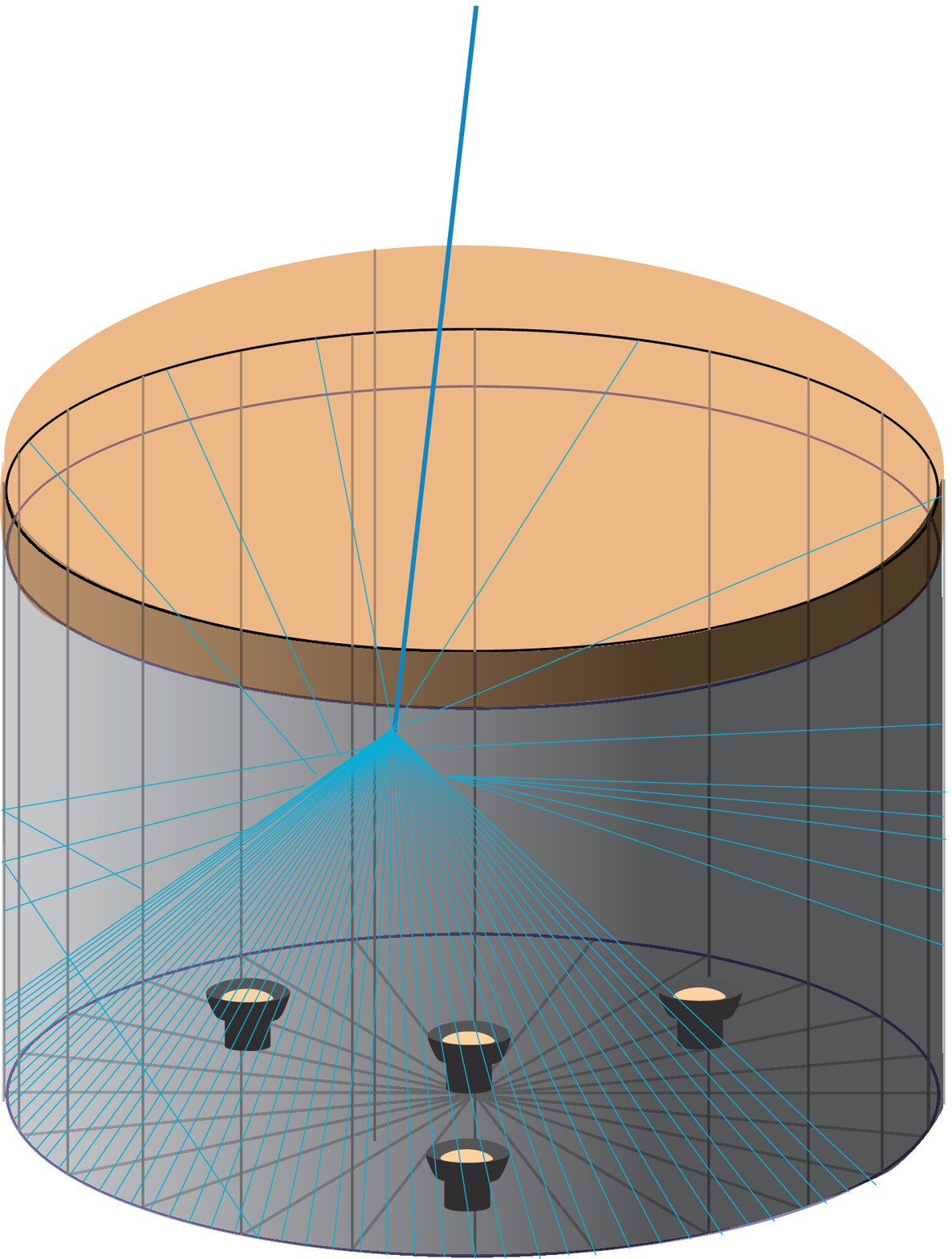}
\includegraphics[width=0.48\textwidth]{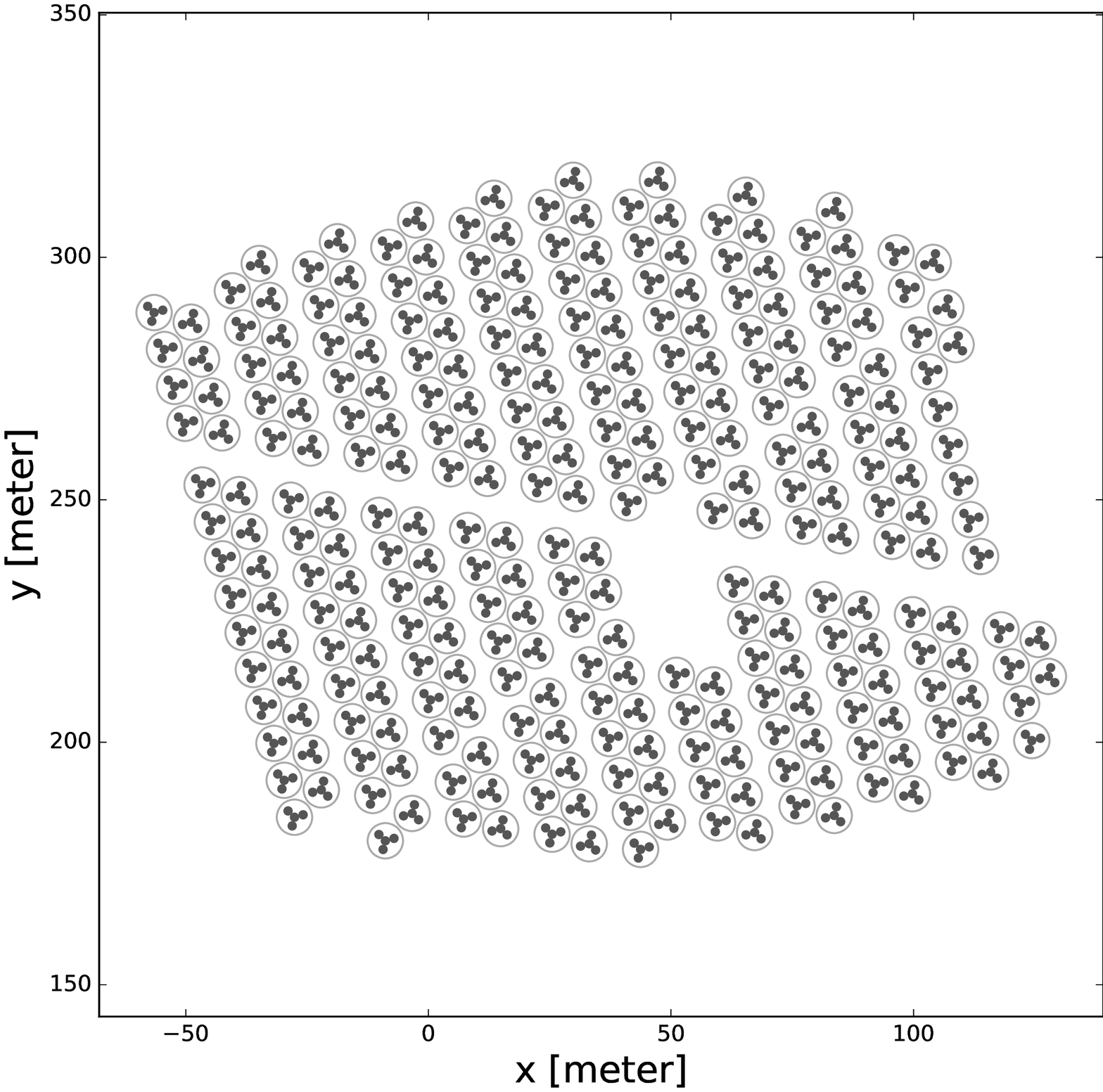}
\caption{The left panel shows a schematic of a single 
HAWC WCD including the steel tank, the covering roof, the
three 8-inch Hamamatsu R5912 PMTs, and one 10-inch Hamamatsu
R7081-MOD
PMT. The tanks are 7.5 meters
diameter and 5 meters high. Water is filled to a depth of 4.5 meters
with 4.0 meters of water above each PMT. 
The right panel shows the layout of the 
completed HAWC instrument, covering 22,000 $\rm{m}^2$.
The location of each WCD is indicated by a large circle 
and PMTs are indicated
with smaller circles. The gap in the center hosts a building
with the data acquisition system.
}
\label{fig:layout}
\end{figure}

The High Altitude Water Cherenkov (HAWC) observatory is a new instrument
sensitive to multi-TeV hadron and gamma-ray air showers,
operating at latitude of +19$^\circ$N at an altitude of 4,100 meters
in the Sierra Negra, Mexico. 
HAWC consists of a large 22,000 $\rm{m}^2$ area densely 
covered
with 300 Water Cherenkov Detectors (WCDs), of which 294 have been 
instrumented. 
Each WCD consists of a 7.3-meter diameter, 5-meter tall steel tank
lined with a plastic bladder and filled with purified water. 
Figure \ref{fig:layout} shows a schematic of the WCD and an overhead
view of the full instrument.
At the bottom of each WCD,
three  
8-inch Hamamatsu R5912 
photomultiplier tubes (PMTs) are anchored in an equilateral
triangle of side length 3.2 meters, with one 10-inch high-quantum efficiency
Hamamatsu R7081
PMT anchored at the center.

A high-energy photon impinging on the atmosphere above HAWC initiates
an extensive electromagnetic air shower. The resulting mix of relativistic 
electrons, positrons
and gamma rays propagates to the ground in a thin tortilla 
of particles at nearly the speed of light. 
Energetic particles that reach the instrument
can interact in the water and produce optical light 
via Cherenkov radiation.
The high altitude of HAWC 
sets the scale for the photon energy that can be detected.
At HAWC's altitude, the shower from a 1 TeV photon from 
directly overhead 
will have about 7\% of the original photon energy left when 
the shower reaches the ground. The
fraction of energy reaching the ground rises to $\sim$28\% at
100 TeV. 
The detector is fully efficient
to gamma rays with a primary energy above $\sim$1 TeV. Lower-energy
photons can be detected when they fluctuate to interact deeper
in the atmosphere than typical. 

The voltages on the HAWC PMTs are chosen to match the PMT gains across
the array.
PMT pulses
are amplified, shaped, and passed through two discriminators at
approximately 1/4 and 4 PEs \citep{hawc111bsl} and digitized. 
The length of time that PMT pulses spend
above these thresholds (time-over-threshold or ToT)
is used to estimate the total amount of charge
collected in the PMT.
Noise arises from a number of sources including PMT afterpulsing,
fragments of sub-threshold air showers, PMT dark noise, and other sources.
The 8-inch PMTs have a hit rate (a hit being each time the PMT signal 
crosses the 1/4 PE threshold) 
due to the combined effect
of these sources 
of 20--30 kHz and the 10-inch PMTs
have a hit rate of 40--50 kHz.

The data from the front-end electronics
is digitized with commercial time-to-digital converters (TDCs) and 
passed to a farm of computers for real-time triggering and processing. 
Events are preserved by the computer farm if they pass
the trigger condition: 
a simple multiplicity trigger, 
requiring some number, $N_{\rm{thresh}}$, PMTs hit within
150 ns. Hits 500 ns prior to a trigger and up to 1000 ns 
after a trigger are also saved for reconstruction. 
During the operation of HAWC, $N_{\rm{thresh}}$ has varied between
20--50. The trigger rate at the
time of writing, 
due primarily to hadronic cosmic-ray
air showers,
is $\sim$24 kHz with $N_{\rm{thresh}}=28$.

The reconstruction process involves determining the direction,
the likelihood for the event to be a photon, and the event's 
size. A first-look reconstruction is applied at the HAWC site. 
In this analysis, 
all the data has been reconstructed again (the fourth revision, or
Pass 4, of
the reconstruction process) off-site in order to have
a uniform dataset and the best calibrations available. 
The chief background to gamma-ray observation is the abundant
hadronic cosmic-ray population. Individual gamma-ray-induced air showers
can be distinguished from cosmic-ray showers by their topology and
the presence of 
deeply penetrating particles at the ground.

The strength of HAWC over the IACT 
technique is that photon showers may be detected
across the entire $\sim$ 2 sr field-of-view of the instrument, day or night,
regardless of weather conditions. As such, HAWC is uniquely suited
to study the long-duration light curve of objects and to 
search for flaring sources in real time. Additionally, since 
sources are observed on every transit, HAWC
obtains thousands of hours of exposure on each source, greatly improving
the sensitivity to the highest-energy photons.

Section \ref{sec:reconstruction} outlines the algorithms by which the
direction, size, and type (photon or hadron) of each shower is determined. 
Section \ref{sec:crabsignal} describes the identification of the gamma-ray
signal from the Crab Nebula. The fit to the Crab energy spectrum, including a
treatment of systematic errors, is
described in Section \ref{sec:spectralfit}. Finally, a discussion of 
the result is presented in Section \ref{sec:discussion}, including 
a comparison to prior spectra measured by peer experiments and 
a computation of the sensitivity of the HAWC instrument, anchored
in the agreement of the HAWC measurement to other experiments.  

\section{Air Shower Reconstruction}
\label{sec:reconstruction}

Events from the detector are reconstructed to determine the arrival 
direction
of the primary particle and the size of the resulting air shower
on the ground, a proxy for the 
primary particle's energy. 
Table \ref{tab:recosteps} summarizes the steps in reconstruction of 
HAWC events as explained below. 

\begin{figure}
\gridline{\fig{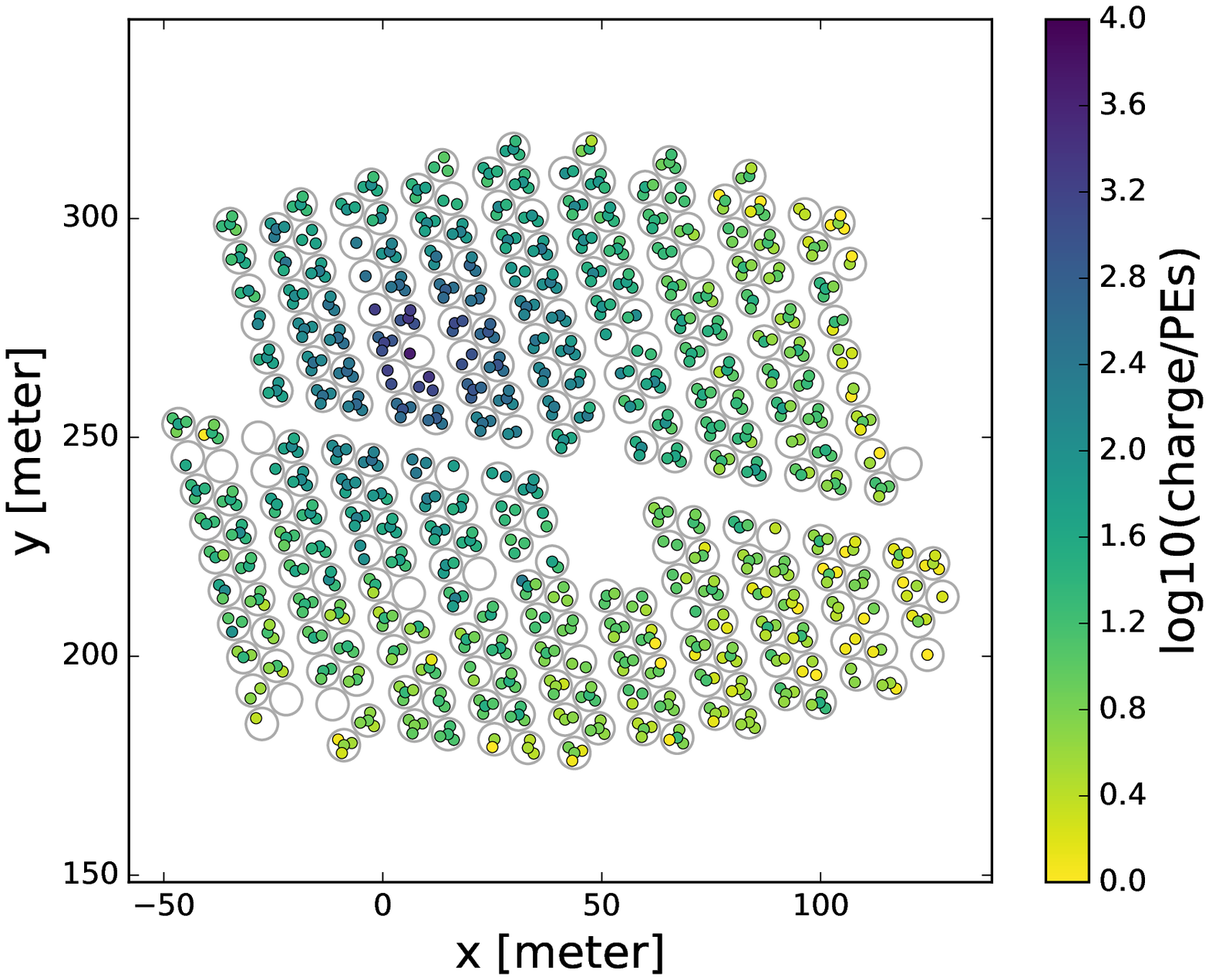}{0.45\textwidth}{(a) Recorded Effective Charge}
          \fig{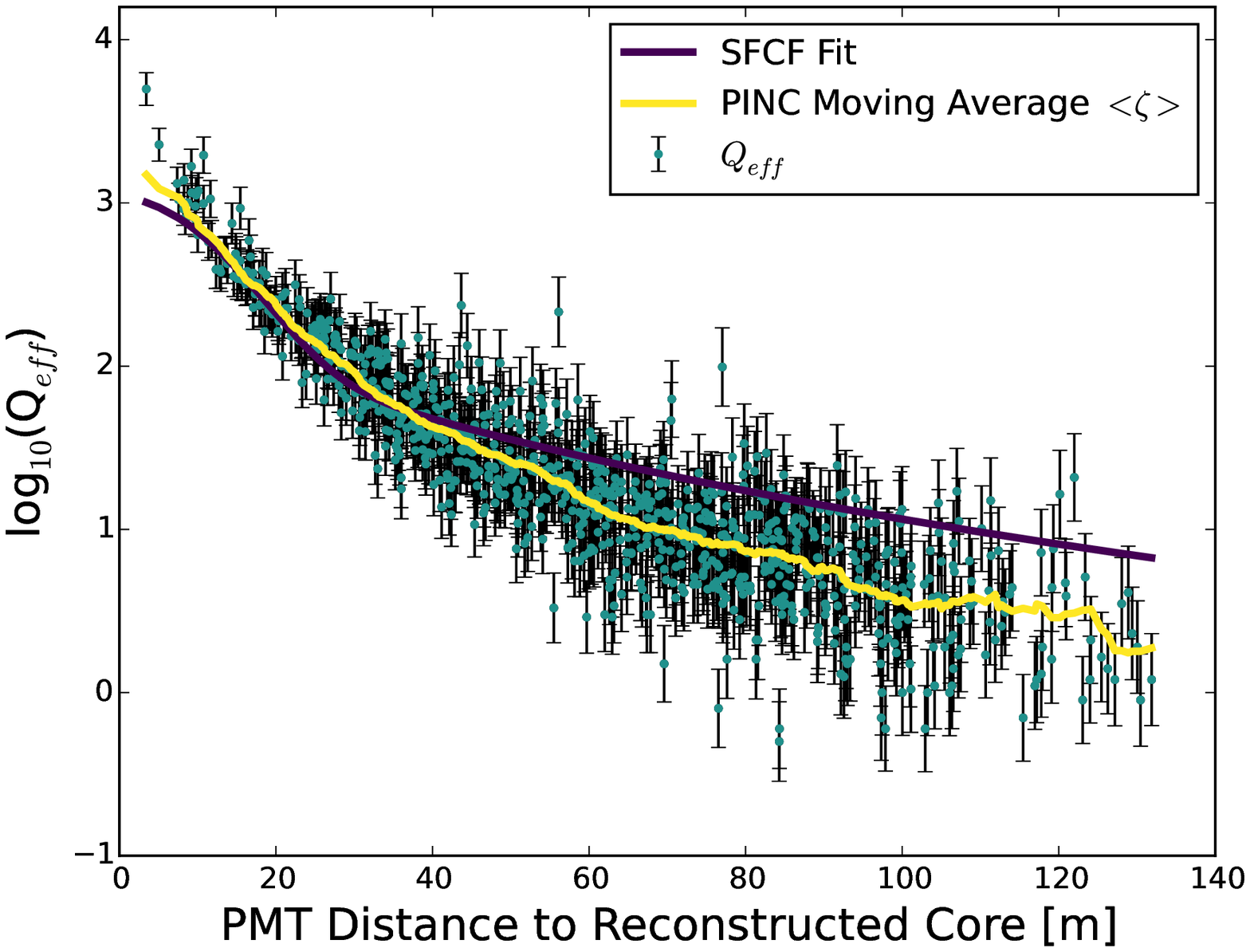}{0.45\textwidth}{(b) Lateral Distribution Function}
          }
\gridline{\fig{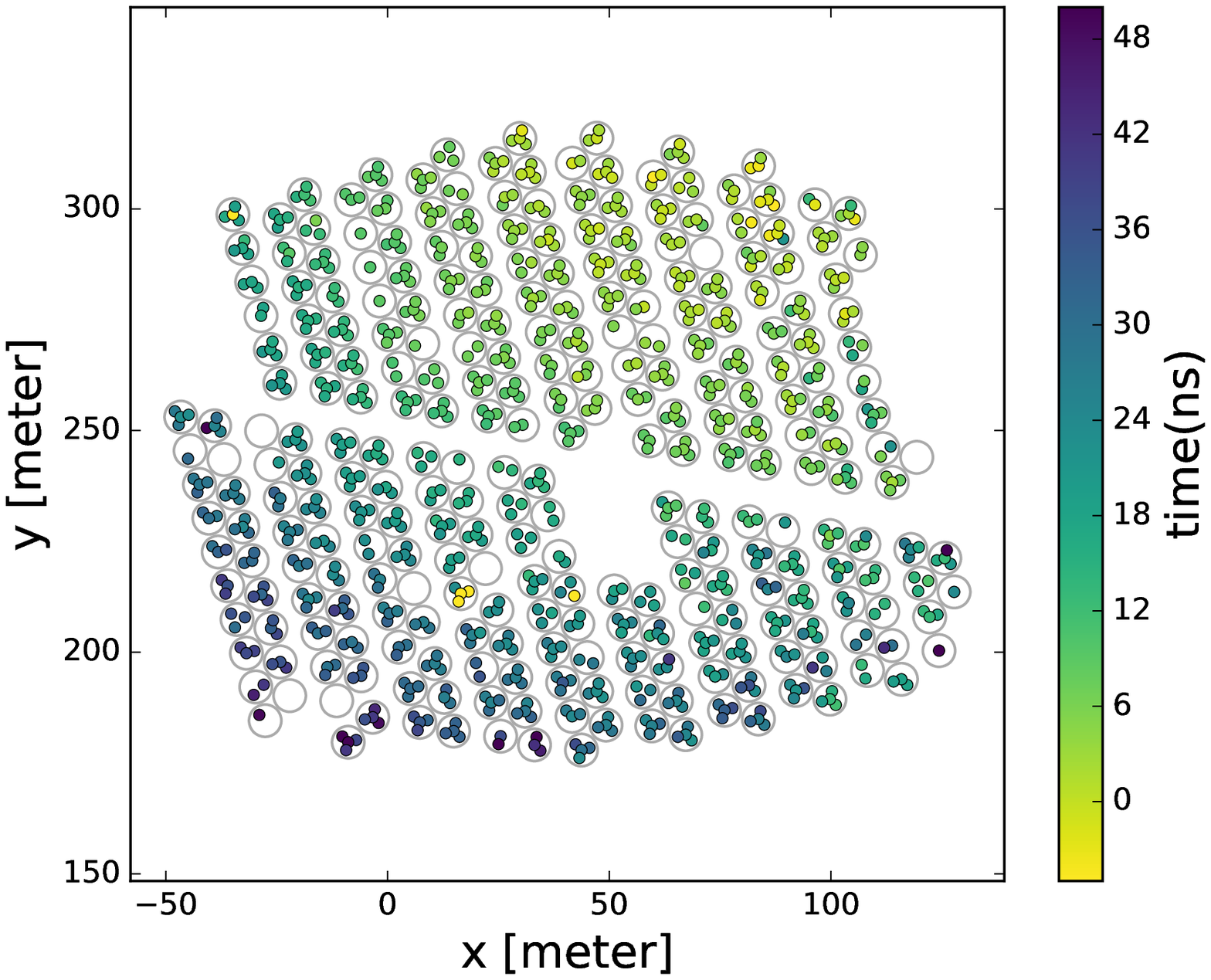}{0.45\textwidth}{(c) Recorded Time}
          \fig{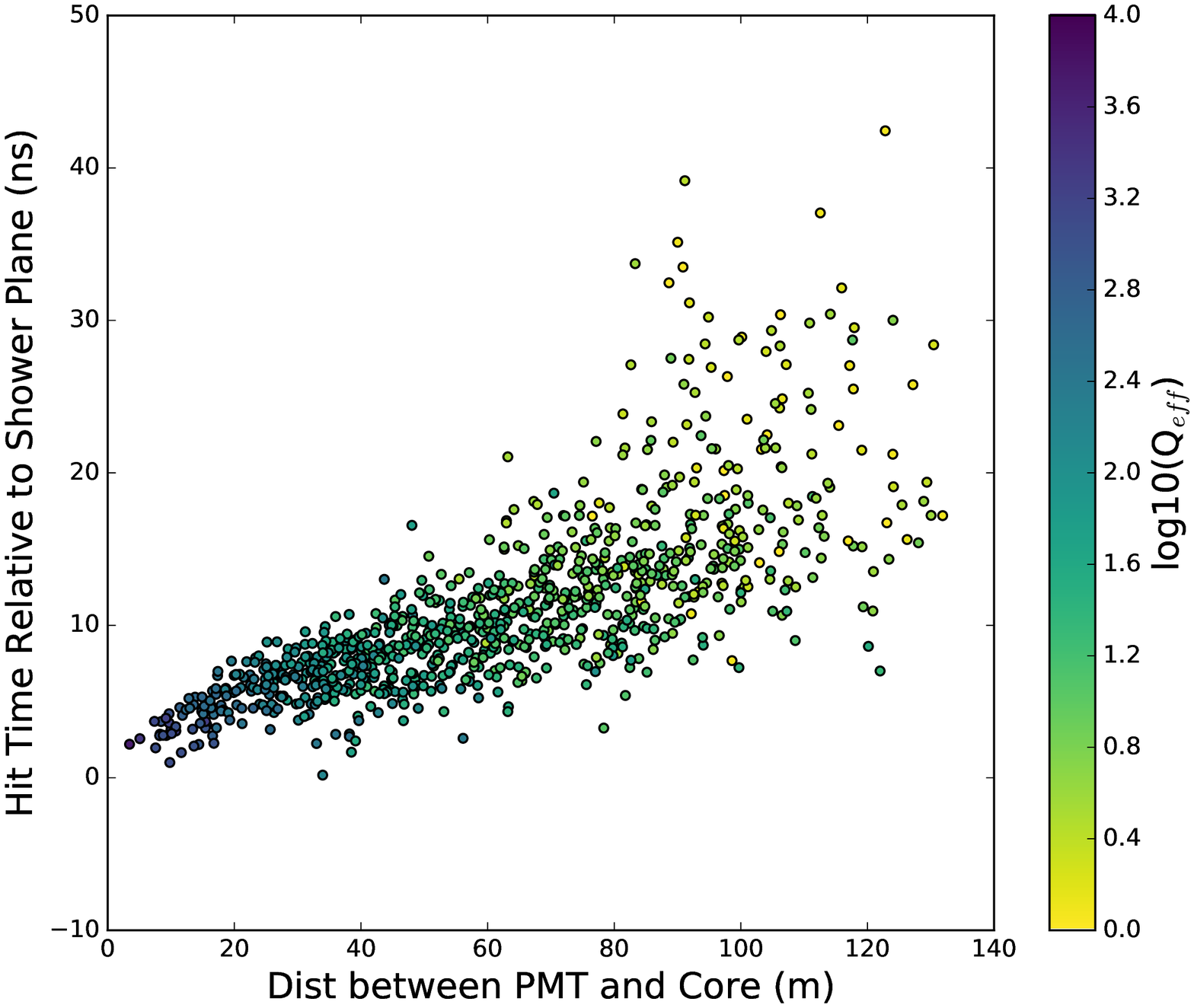}{0.45\textwidth}{(d) Shower Curvature/Sampling}
          }
\caption{These figures illustrate a high-confidence
gamma ray from the Crab Nebula, taken from a dataset with a better than
10:1 signal to background ratio. Panels a and c show an overhead view
of the HAWC instrument with large circles to indicate individual
WCDs and small circles to indicate individual PMTs. Sporadic PMTs
 have been removed as described in Section \ref{sec:size}. 
Panel a shows
the PMT Effective Charge, defined in Section \ref{sec:calib}, 
for each PMT that recorded a hit during the
event. The shower core is evident.  
Panel c shows the time each PMT recorded a hit with a color pattern
due to the inclination of the shower. Panel b shows the
lateral distribution function, the effective charge recorded as a function
of the distance from the hit PMT to the reconstructed core. 
The fitted function from the core fit (Section \ref{sec:corereconstruction})
is overlaid. From this 
distribution, the photon/hadron separation parameters $\mathcal{C}$ and
$\mathcal{P}$ are computed (Section \ref{sec:ghsep}) and the 
moving average used in the computation of $\mathcal{P}$ is shown. 
Finally, Panel d
shows the time each PMT recorded a hit 
relative to a perfect shower plane (under the
assumption that the photon came from the Crab,as explained in Section \ref{sec:directionreconstruction}) as a function of the distance
of the hit from the shower core. The need for a timing correction before
the plane fit (due
to the curvature and sampling effects) 
is evident in Panel d.}
\label{fig:sampleevent}
\end{figure}

To illustrate the event reconstruction, 
Figure \ref{fig:sampleevent} shows a strong gamma-ray candidate from the 
Crab Nebula.
In Section \ref{sec:simulation}, the simulation is
briefly described as it is key to evaluating the reconstruction process.
Section \ref{sec:calib} describes the calibration, by which the time and
light
level in individual PMTs are determined. Section \ref{sec:size} discusses
the selection of PMT signals for reconstruction and the
event size measurement. The direction reconstruction occurs in two 
steps, first the core reconstruction, 
described in Section \ref{sec:corereconstruction}, and then the 
direction determination, 
described in Section \ref{sec:directionreconstruction}. 
The air shower
core, the dense concentration of particles along the direction 
of the original primary, is needed to make the best reconstruction
of the air shower's direction since the air shower arrival front
is delayed from a pure plane, depending on the distance from the
core.
The identification
of photon candidates is presented in Section \ref{sec:ghsep}. The directional
fit is iterated to suppress noise and this iteration is explained
in Section \ref{sec:refinement}.

\begin{table}
\begin{tabular}{ c l l }
Step & Description & Hit Selection \\
\hline
1 & Calibration &  \\
2 & Hit Selection &   \\
3 & Center-of-Mass Core Reconstruction &  Selected Hits \\
4 & SFCF Core - First Pass  &  Selected Hits\\
5 & Direction  - First Pass & Selected Hits\\
6 & SFCF Core - Second Pass  &  Selected Hits within 50 ns of First-pass Plane\\
7 & Direction - Second Pass&  Selected Hits within 50 ns of First-pass Plane\\
8 & Compactness & Selected Hits within 20 ns of Second-pass Plane\\
9 & PINCness & Selected Hits within 20 ns of Second-pass Plane\\
\end{tabular}
\caption{Steps in the HAWC Event Reconstruction. The best core and direction
reconstructions are applied with gradual narrowing of the hits used.}
\label{tab:recosteps}
\end{table}

\subsection{Simulation}
\label{sec:simulation}

The HAWC instrument is modeled using a combination of community-standard 
simulation
packages and custom software. 
The CORSIKA package (v7.4000) is used for simulation of air showers,
propagating the primary particles through the atmosphere to
the ground \citep{corsika}.
At ground level, a GEANT 4 
simulation 
(v4.10.00)
of the shower particles is used
to propagate the ground-level particles through the HAWC tanks
and to track the Cherenkov photons to the faces of the PMTs
\citep{geant4}. 

The response of the PMTs and the calibration are approximated with a custom
simulation that assumes that recorded light is faithfully detected with some 
efficiency and an uncertainty in the logarithm of the total charge
recorded. 
Decorrelated single PE noise is added. 
The absolute 
PMT efficiency for detecting Cherenkov photons is established by scaling 
the simulated PMT response to vertical muons to match data. 
Most muons passing through HAWC are minimum ionizing with nearly
constant energy loss. Vertical muons, therefore, are a nearly
constant light source and convenient for establishing to total
PMT efficiency.
Simulated
events are subsequently 
reconstructed by the same procedure as experimental data
to study the performance of the algorithms.

\subsection{Calibration}
\label{sec:calib}

The first step in the reconstruction process is 
calibration, the processes by which true time 
and light level 
in each PMT are estimated from the TDC-measured threshold-crossing
times of each PMT \citep{calibrationicrc2013,calibrationicrc2015}. 

The calibration associates the measured ToT in each PMT
with the true number of PEs. 
To give a sense of scale, the ToT for a single PE crossing the low-threshold
discriminator (about 1/4 PE) is $\sim$100 ns. Above a few PEs, the 
higher-threshold (about 4 PEs) ToT is used for charge assignment
and a high-threshold ToT of 400 ns roughly 
corresponds to a charge of $10^4$ PEs. The time scale for
these ToTs is determined by the shaping of the front-end electronics
and is chosen to be longer than the characteristic 
arrival time distribution of PEs during an air shower so as to 
integrate the whole air shower arrival into one PMT hit. 

In addition to the PE measurement, the calibration system accounts for
electronic slewing of the PMT waveform: Lower-PE waveforms cross
threshold later than contemporaneous high-PE waveforms. 

Subsequent reconstruction 
algorithms treat all PMTs, the 8-inch and 10-inch,
as identical, despite the larger size and greater efficiency of the
10-inch PMTs. To accommodate this, an effective charge 
$Q_{\rm{eff}}$ is defined. For $Q_{\rm{eff}}$, PE values from the central 
10-inch PMTs are scaled by a
factor of 0.46 to place them {\it{on par}} with the 8-inch PMTs.

Finally, each PMT has a single calibrated timing
offset that accounts for the different cable lengths and 
any other timing delays that may differ from PMT to PMT. These
delays are measured to within a few ns by the calibration system
and are refined to sub-ns precision by iterated fits to 
hadronic air showers. Since the hadronic background is isotropic,
the point of maximum cosmic-ray density is overhead and the PMT
timing pedestals are chosen to ensure this is true. A final
small ($\sim$0.2$^\circ$) rotation of all events is performed
to ensure that the Crab Nebula appears in its known location.

Throughout the analysis, the Crab is assumed to be at a location
of 83.63$^\circ$ right ascension and 22.01$^\circ$ declination, in the
J2000.0 epoch, taken
from \cite{hegracrab}. While the pulsar position is 
known more precisely (e.g. \cite {oldcrabpulsarradio}), this precision
is sufficient for use in HAWC.

\subsection{Hit Selection and Event Size Bins}
\label{sec:size}

As described in Section \ref{sec:intro}, the HAWC DAQ 
records 1.5 $\mu$s of data from all PMTs that have a hit 
during an air shower event. A subset of these
hits are selected for the air shower fit. To be used for the air shower fit,
hits must be found between -150 and +400 ns around the trigger time. Hits
are removed if they occur shortly after a high-charge hit under the
assumption that these hits are likely contaminated with 
afterpulses. Additionally, hits are
removed if they have a pattern of TDC crossings that is not characteristic
of real light; they cannot be calibrated accurately. Finally, each
channel has an individual maximum calibrated charge, typically
a few thousand PEs, but no more than 10$^4$ PEs, above which the PMTs
are not used. Above $\sim$10$^4$ PEs, corresponding to a ToT of $\sim$400
ns, prompt afterpulsing in the PMTs can artificially lengthen the
ToT measurement giving a false measurement. 
Channels
are considered available for reconstruction if they have a live 
PMT taking data which has not been removed by one of these cuts.

The angular error and the ability to distinguish photon events
from hadron events is strongly dependent on the energy and size
of events on the ground. 
We adopt analysis cuts and
an angular resolution description that depends on this measured size.
The data is divided into 9 size bins, $\mathcal{B}$,
as outlined in Table \ref{table:cuts}. The size of the event is defined
as the ratio of the number of PMT hits used by the event reconstruction
to the total number of PMTs available for reconstruction, $f_{\rm{hit}}$. This definition
allows for relative stability of the binning when PMTs are
occasionally taken out of service.

For this analysis, events are only used if they have more than 6.7\%
of the available PMTs seeing light. Since typically 1000 PMTs are
available, typically a minimum of 70 PMTs is needed for an event. 
This is
substantially higher than the trigger threshold. The data between
the trigger threshold and the threshold for $\mathcal{B}=1$ in this analysis
consists of real air showers, and techniques to recover these events
and lower the energy threshold, beyond what is presented here, are under
study.

\begin{table}
\begin{tabular}{ c c c c c c}
  $\mathcal{B}$ &  $f_{\rm{hit}}$ & $\psi_{68}$ & $\mathcal{P}$ Maximum & $\mathcal{C}$ Minimum & Crab Excess Per Transit\\
\hline 
  1 &  6.7 - 10.5\%   &  1.03                 & $<$2.2  &  $>$7.0   & 68.4 $\pm$ 5.0  \\
  2 & 10.5 - 16.2\%   & 0.69                  & 3.0  &  9.0   & 51.7 $\pm$ 1.9  \\
  3 & 16.2 - 24.7\%   & 0.50                  & 2.3  &  11.0  &  27.9 $\pm$ 0.8 \\
  4 & 24.7 - 35.6\%   & 0.39                  & 1.9  &  15.0  &  10.58 $\pm$ 0.26 \\
  5 & 35.6 - 48.5\%   & 0.30                  & 1.9  &  18.0  &  4.62 $\pm$ 0.13 \\
  6 & 48.5 - 61.8\%   & 0.28                  & 1.7  &  17.0  &  1.783 $\pm$ 0.072 \\
  7 & 61.8 - 74.0\%   & 0.22                  & 1.8  &  15.0  &  1.024 $\pm$ 0.053 \\
  8 & 74.0 - 84.0\%   & 0.20                  & 1.8  &  15.0  &  0.433 $\pm$ 0.033 \\
  9 & 84.0 - 100.0\%  & 0.17                  & 1.6  &  3.0   &  0.407 $\pm$ 0.032 \\
\end{tabular}
\caption{Cuts used for the analysis. The definition of the size bin $\mathcal{B}$
is given by the fraction of available PMTs, $f_{\rm{hit}}$, 
that record light during the 
event. Larger events are reconstructed better and $\psi_{68}$, the 
angular bin that contains 68\% of the events, reduces dramatically
for larger events. 
The parameters $\mathcal{P}$ and $\mathcal{C}$ (Section \ref{sec:ghsep})
characterize the 
charge topology and are used to remove hadronic air shower events. 
Events with a $\mathcal{P}$ less than indicated
and a $\mathcal{C}$ greater than indicated are considered
photon candidates. The cuts are established by optimizing the statistical
significance of the Crab and trend toward harder cuts at larger size events. The number
of excess events from the Crab in each $\mathcal{B}$ bin per transit 
is shown as well. }
\label{table:cuts}
\end{table}

\begin{figure}[h]
\centering
\includegraphics[width=0.65\textwidth]{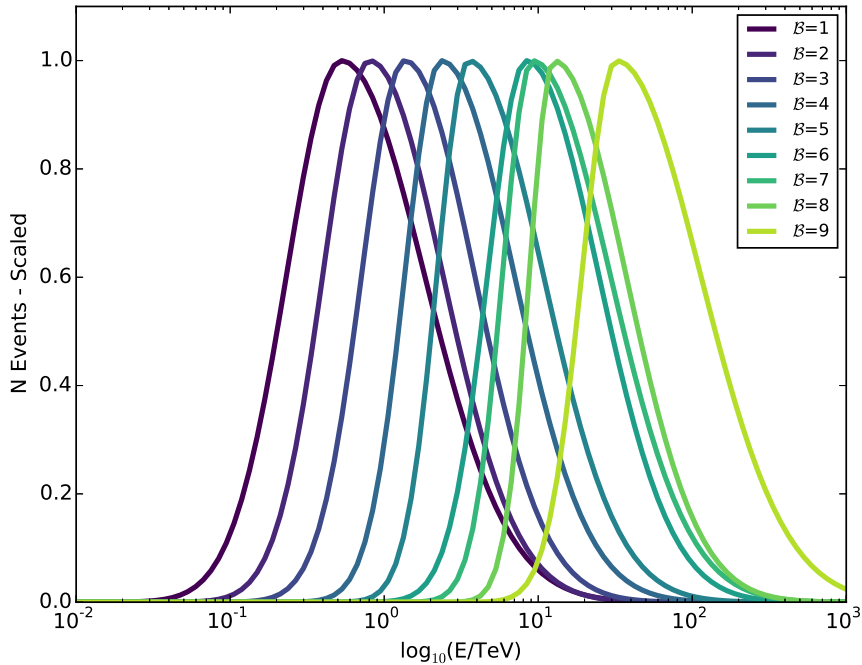}
\caption{Fits to the true energy distribution of photons from a source 
with a spectrum
of the form $E^{-2.63}$ at a declination of +20$^\circ$N for $\mathcal{B}$ 
between 1 and 9, summed across a transit of the source. 
Better energy resolution and dynamic range can 
be achieved with a more sophisticated variable that takes into account the 
zenith angle of events and the total light level on the ground.  
The curves have been scaled to the same vertical height for display. 
}
\label{fig:energy}
\end{figure}

Figure \ref{fig:energy} shows the distribution of true energies
as a function of the $\mathcal{B}$ of the events. The distribution
of energies naturally depends heavily on the source itself, both 
its spectrum and the angle at which it culminates during its transit.
A
pure power-law spectrum with a shape of $E^{-2.63}$ and a declination 
of 20$^\circ$ was assumed for 
this figure. As $\mathcal{B}$ is a simple variable ---
containing no correction for zenith angle,
impact position, or light level in the event --- the energy distribution
of $\mathcal{B}$ bins is wide. Section \ref{sec:improvements} discusses
planned
improvements to this event parameter that will measure the energy of 
astrophysical gamma rays better.

Bin $\mathcal{B}=9$ bears particular attention. It is an ``overflow'' bin
containing events which have between
84\% and 100\% of the PMTs in the detector seeing light. Typically, a
10 TeV photon will hit nearly every sensor
and the $\mathcal{B}$ variable has no dynamic range above this energy.
This limit is not intrinsic to HAWC and variables that utilize the light
level seen in PMTs on the ground, similar to what was used
in the original sensitivity study \citep{sensipaper},
have dynamic range above 100 TeV. These variables, not used in this 
analysis, will improve the identification of high-energy events. This is
discussed farther in Section \ref{sec:improvements}.

\subsection{Core Reconstruction}
\label{sec:corereconstruction}

In an air shower, the concentration of secondary particles is highest
along the trajectory of the original primary particle, termed the air
shower core.
Determining the position of the core on the ground
is key to reconstructing the direction
of the primary particle.
In the sample event, Figure \ref{fig:sampleevent}, the air shower
core is evident in Figure \ref{fig:sampleevent}a. The image is 
an overhead view of the HAWC detector with circles indicating the WCD
location and the PMTs within the WCDs. The colors indicate the amount of
light (measured in units of PEs) seen in each PMT.
The air shower core is evident as the point of maximum PE density.

The PE distribution on the ground is fit
with a function that decreases monotonically with the distance
from the shower core. The signal in the $i$th PMT, $S_i$, is presumed
to be

\begin{equation}
S_i = S(A, \vec{x}, \vec{x}_i) = A \Big(\frac{1}{2\pi \sigma^2}e^{-{|\vec{x}_i - \vec{x}|^2}/{2\sigma^2}} + \frac{N}{(0.5 + {|\vec{x}_i - \vec{x}|} /{R_{m}})^3}\Big)
\label{eqn:sfcf}
\end{equation}

\noindent where $\vec{x}$ is the core location, $\vec{x}_i$ is the location 
of the measurement, $R_{m}$ is the Moli\`{e}re radius of the atmosphere,
approximately 120 m at HAWC altitude, 
$\sigma$ is the width of the Gaussian,
and $N$ is the normalization of the tail.  
Fixed values of $\sigma=10$ m and $N=5 \cdot 10^{-5}$ are used.
This leaves three free
parameters, the core location and overall amplitude, $A$. 

The functional form used in this algorithm, 
termed the Super Fast Core Fit (SFCF), is a simplification
of a modified Nishimura-Kamata-Greisen (NKG) 
function \citep{nkgfunction} and is chosen for 
rapid fitting of air shower cores. 
The NKG function
has an additional free parameter, the shower age, 
and involves computationally intensive power law and 
gamma function evaluation. The 
SFCF hypothesis in Equation \ref{eqn:sfcf} is similar but numerical
minimization can converge faster because: the function is simpler,
the derivatives
are computed analytically, and the lack of a pole at the core location.

Figure \ref{fig:sampleevent}b 
shows the recorded charge in each
PMT as a function of the PMT's distance along the ground
to the reconstructed shower core.
The fit for this event is shown along with the PINCness moving average
from Section \ref{sec:ghsep}. While the full NKG function 
would describe the lateral distribution better, 
the SFCF form allows rapid
identification the center of showers and this is sufficient for the present
analysis. Cores can be localized to a median error of $\sim$2 meters
for large events ($\mathcal{B}=8$) and $\sim$4 meters for small
events ($\mathcal{B}=3$) for gamma-ray 
events with a core that lands on the 
HAWC detector. 
The error in reconstructing the shower core increases as the core
moves further from the array. For example, a shower with a core
that is 50 meters from the edge of the array will have an error in 
the location of the core of $\sim$35 meters.

\subsection{Direction Reconstruction}
\label{sec:directionreconstruction}

To first order, the air shower particles arrive on a plane 
defined by the speed of light and direction of the primary particle.
In fact, the shower front has a slight conical shape centered at 
the air shower core.
Several effects lead to this shape.
First, particles far from the core arrive due to multiple scattering and
longer travel distances. Second, the multiple scattering of particles
at the edges of the shower leads to a broader arrival time distribution
than at the core. Since the number of particles decreases with increasing 
distance from the shower core there are fewer opportunities to sample 
from the particle arrival time distribution. This decrease in sampling
leads to a delay in the measured arrival time of the shower. 

The conical shape of the air shower front can be easily seen with the sample
event in Figure \ref{fig:sampleevent}. 
Figure \ref{fig:sampleevent}c shows, for each PMT in the sample event,
the calibrated time each PMT saw light. The 
color trend is due to the inclined direction of the air shower.
Taking out this inclination, we can see the curved shower front. 
The event in Figure \ref{fig:sampleevent} was chosen for display because 
--- taken
from a high signal-to-background sample of data --- it 
is very likely a photon from the Crab. 
If we assert the origin of this particle is the Crab Nebula, we know
the air shower plane precisely and
can make Figure \ref{fig:sampleevent}d: We adjust the times of each PMT
hit assuming a pure planar air shower originating at the Crab Nebula. 
Figure \ref{fig:sampleevent}d
shows the plane-corrected time as a function of the PMT distance from the
core of the air shower along the ground. A pure planar shower would
be a horizontal line on this figure. The delay of particles far from
the core, is evident. 

In the present analysis, we use the reconstructed core location 
to correct for this effect. 
A combined curvature/sampling correction --- a function
of the distance of hits from the shower core and the total charge recorded
in the PMT --- is utilized for this correction. 
The curvature/sampling correction is based on a combination
of simulation and Crab observations. The rough functional form is tabulated
using gamma-ray simulation. The simulation-optimized 
curvature/sampling correction
yields a measured angular resolution approximately
a factor of 2 worse than predicted from simulation. The origin of this 
discrepancy is likely due to some oversimplification
of the electronics simulation. Repeated fits to the Crab have yielded
a modification to the
curvature/sampling correction that is a simple quadratic function of
the distance between a hit and the shower core. While the origin 
of the discrepancy is under investigation, it amounts to a relatively
small correction, approximately 2 ns/100 meters. Nonetheless, the improvement
in the angular resolution is nearly a factor of two for all $\mathcal{B}$.
Remaining disagreement between the simulated and measured angular resolution
is adopted as a systematic error.

After correcting for the sampling and curvature, the angular fit is a 
simple $\chi^2$ planar fit and has been described before \citep{milagrocrab}.

\subsection{Photon/Hadron Separation}
\label{sec:ghsep}

Hadronic cosmic rays are the most abundant particles producing
air showers in HAWC and
constitute the chief background to high-energy photon
observation. The air showers produced by high-energy 
cosmic rays and gamma rays differ: gamma-ray showers are
pure electromagnetic showers with few muons
or pions. Conversely, hadronic cosmic rays produce hadronic 
showers rich with pions, muons, other hadronic secondaries,
and structure due to the showering of daughter particles 
created with high transverse momentum. 
In HAWC, these two types of showers appear quite different,
particularly for showers above several TeV. 

\begin{figure}[h]
\centering
\includegraphics[width=0.45\textwidth]{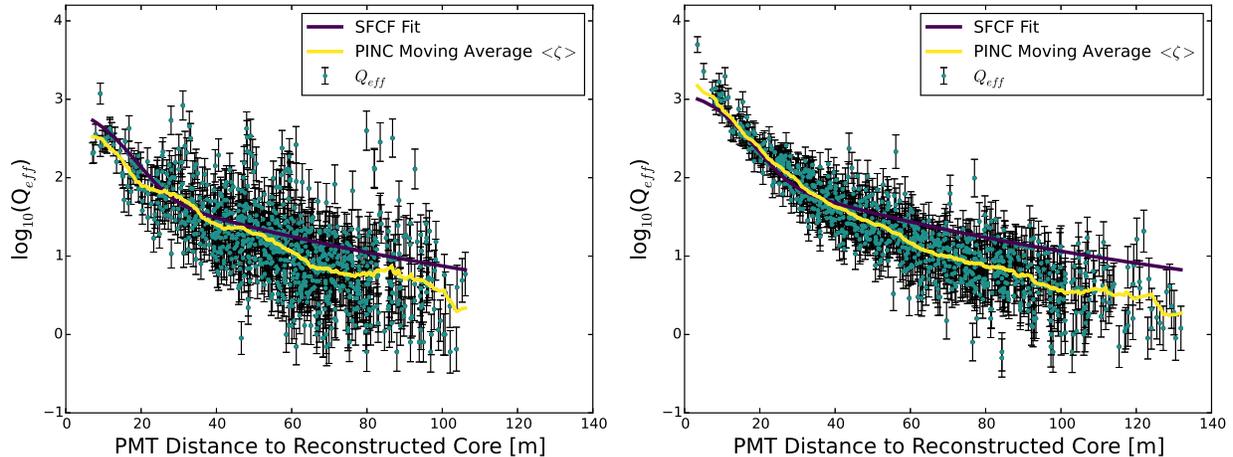}
\includegraphics[width=0.45\textwidth]{pinc.rec.event-4.7606.eps}
\caption{Lateral distribution functions of an obvious cosmic ray (left) and
a photon candidate from the Crab Nebula (right). 
The cosmic ray has isolated high-charge hits far from the shower
core due to penetrating particles in the hadronic air shower. These
features are absent in the gamma-ray shower.
}
\label{fig:ghsepexample}
\end{figure}

Figure \ref{fig:ghsepexample} shows the lateral distributions for two
showers, an obvious cosmic ray (left) and a strong photon candidate (right)
from the 
Crab Nebula. The effective light level 
$Q_{\rm{eff}}$ falls off for hits further from the shower core in both
showers, but in the hadronic shower there are sporadic high-charge
hits far from the air shower's center. This clumpiness is 
characteristic of hadronic showers and arises from a combination of 
penetrating particles (primarily muons) and hadronic sub-showers
which are largely absent in photon-induced showers.

Two parameters are used to identify cosmic-ray events. 
The first parameter, compactness, was used in the 
sensitivity study \citep{sensipaper}.
The variable ${\rm{CxPE}}_{40}$ is the effective 
charge measured in the PMT with the
largest effective charge outside a radius of 40 meters from the shower core.
We then define the compactness, $\mathcal{C}$, as

\begin{equation}
\mathcal{C} = {{\rm{N_{hit}}} \over {\rm{CxPE_{40}}}}
\end{equation}

\noindent where ${\rm{N_{hit}}}$ is the number of hit PMTs during the air shower.
${\rm{CxPE}}_{40}$ is typically large for a hadronic event, so $\mathcal{C}$
is small. 

In addition to the largest hit outside the core, the 
``clumpiness'' of the air shower is quantified 
with a parameter $\mathcal{P}$,
termed the PINCness of an event (short for Parameter for
Identifying Nuclear Cosmic-rays). $\mathcal{P}$ is defined using
the lateral distribution function of the air shower, seen in Figure 
\ref{fig:ghsepexample}.
Each of the PMT hits, $i$, has a measured effective charge $Q_{\rm{eff},i}$.
$\mathcal{P}$ is computed using the logarithm of this charge
$\zeta_{i}={\rm{log}}_{10}(Q_{\rm{eff},i})$. For each hit, an expectation
is assigned $\langle\zeta_i\rangle$ 
by averaging the $\zeta_i$ in all PMTs
contained in an annulus containing the hit, with a width of 5 meters, 
centered at the core of the air shower.

$\mathcal{P}$ is then calculated using the $\chi^2$ formula:

\begin{equation}
\mathcal{P}= {1 \over N} {\sum_{i=0}^{N}  { {(\zeta_i - \langle\zeta_i\rangle)^2} \over{ {\sigma_{\zeta_i}}^2}   }}
\end{equation}

\noindent The errors $\sigma_{\zeta_i}$ are assigned from a study of a sample strong gamma-ray candidates in the vicinity of the Crab.
 
The $\mathcal{P}$ variable essentially requires axial smoothness.
Figure \ref{fig:ghsepexample}
shows the moving average $<\zeta_i>$ for two sample events. The hadronic 
event in Figure \ref{fig:ghsepexample} is ``clumpy'' and
has several hits that differ
sharply from the moving average yielding a large $\mathcal{P}$.

\begin{figure}
\gridline{\fig{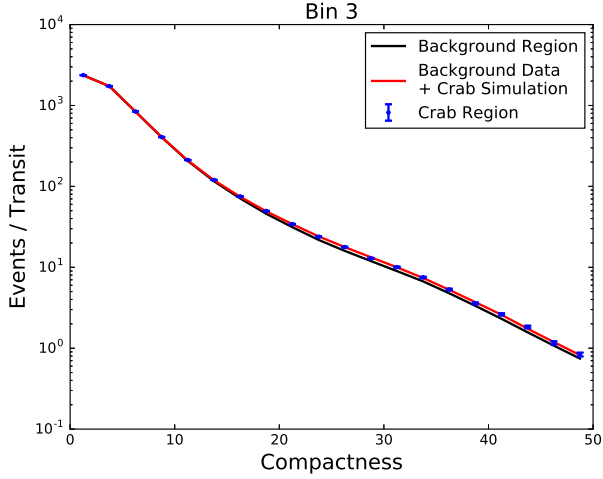}{0.45\textwidth}{(a) $\mathcal{B}=3$ Signal and Background}
          \fig{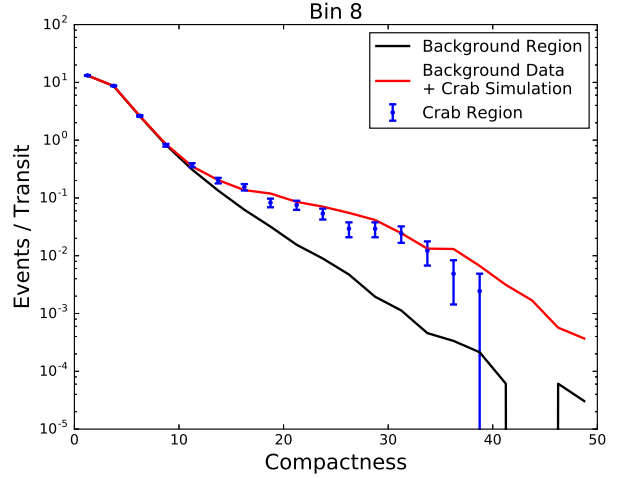}{0.45\textwidth}{(b) $\mathcal{B}=8$ Signal and Background}
          }
\gridline{\fig{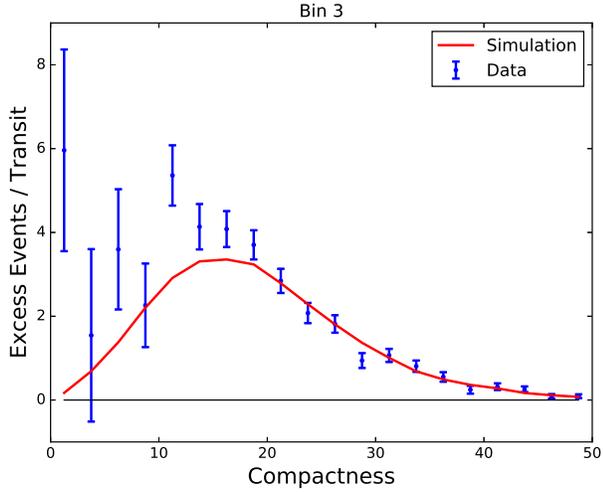}{0.45\textwidth}{(c) $\mathcal{B}=3$ Background Subtracted}
          \fig{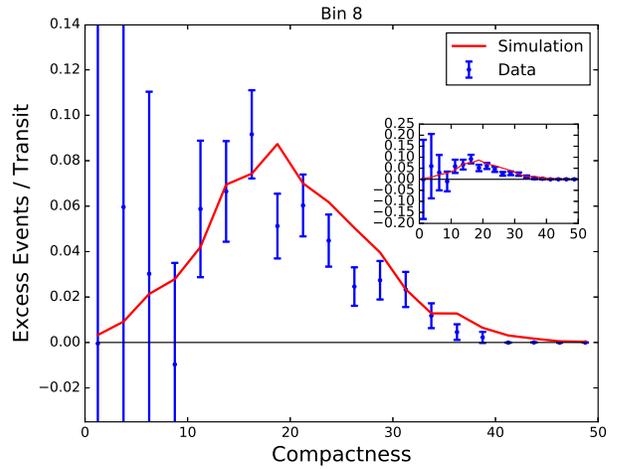}{0.45\textwidth}{(d) $\mathcal{B}=8$ Background Subtracted}
          }
\caption{Compactness, $\mathcal{C}$, distribution for events in the vicinity of the Crab
for $\mathcal{B}$=3 (left panels) and $\mathcal{B}$=8 (right panels).
The top two figures show the raw recorded events in the vicinity of the
Crab itself. 
Events within 1.5$\cdot\psi_{68}$ from Table \ref{table:cuts} are used to 
define
this ``Crab region''. 
A background expectation is generated by scaling the compactness
distribution
from a large annulus around the Crab. 
The bottom figures show the measured distribution from the Crab with this
background level subtracted. Predictions from photon simulation are overlaid
showing that these variables are well-modeled. Larger photon showers 
are typically easier to identify from hadrons
as evidenced in the figures. In Panel d, the error bars in the
background-dominated region are quite large and zoomed to show the 
photon distribution; an inset shows the entire
distribution.}
\label{fig:compactdistribution}
\end{figure}

\begin{figure}
\gridline{\fig{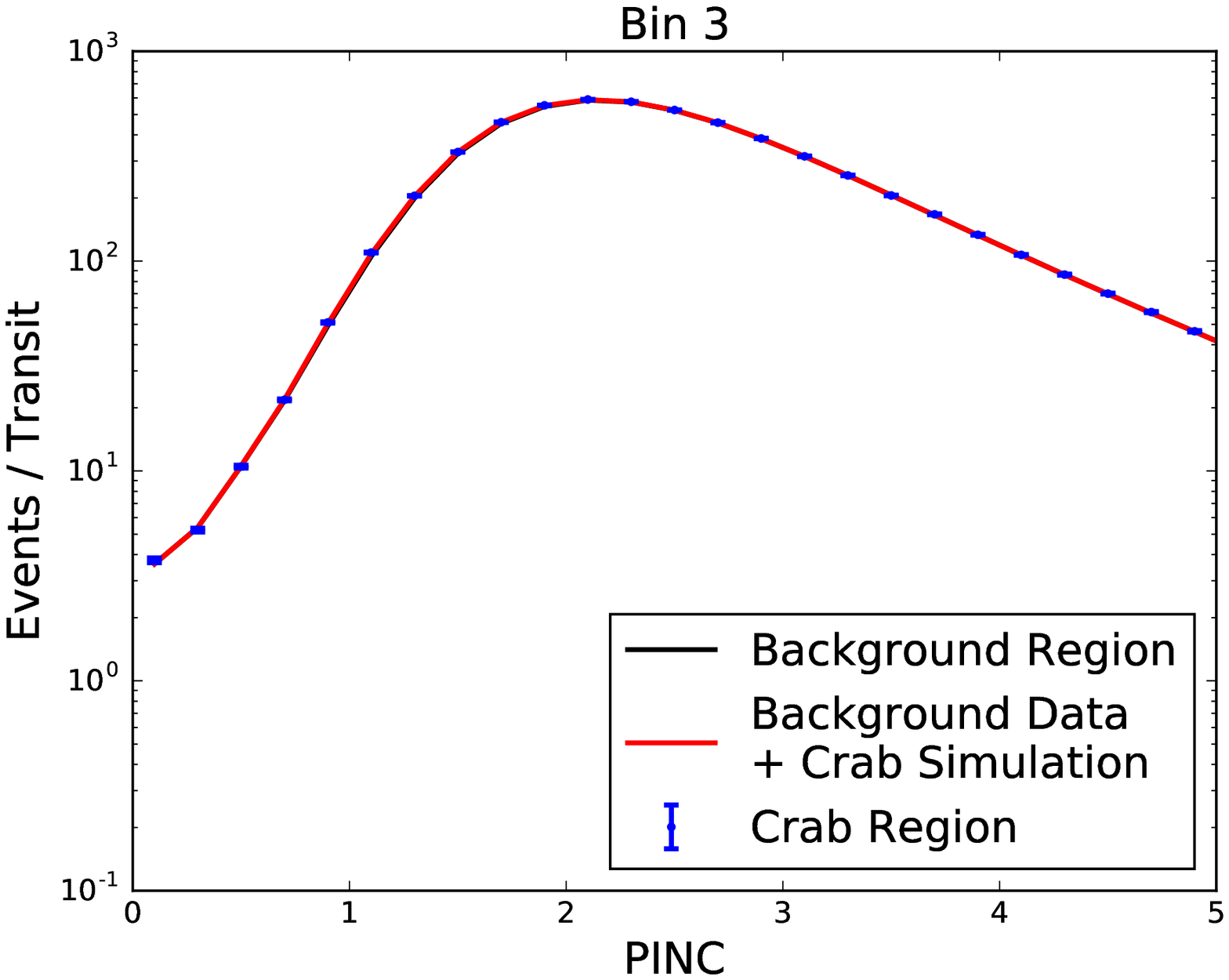}{0.45\textwidth}{(a) $\mathcal{B}=3$ Signal and Background}
          \fig{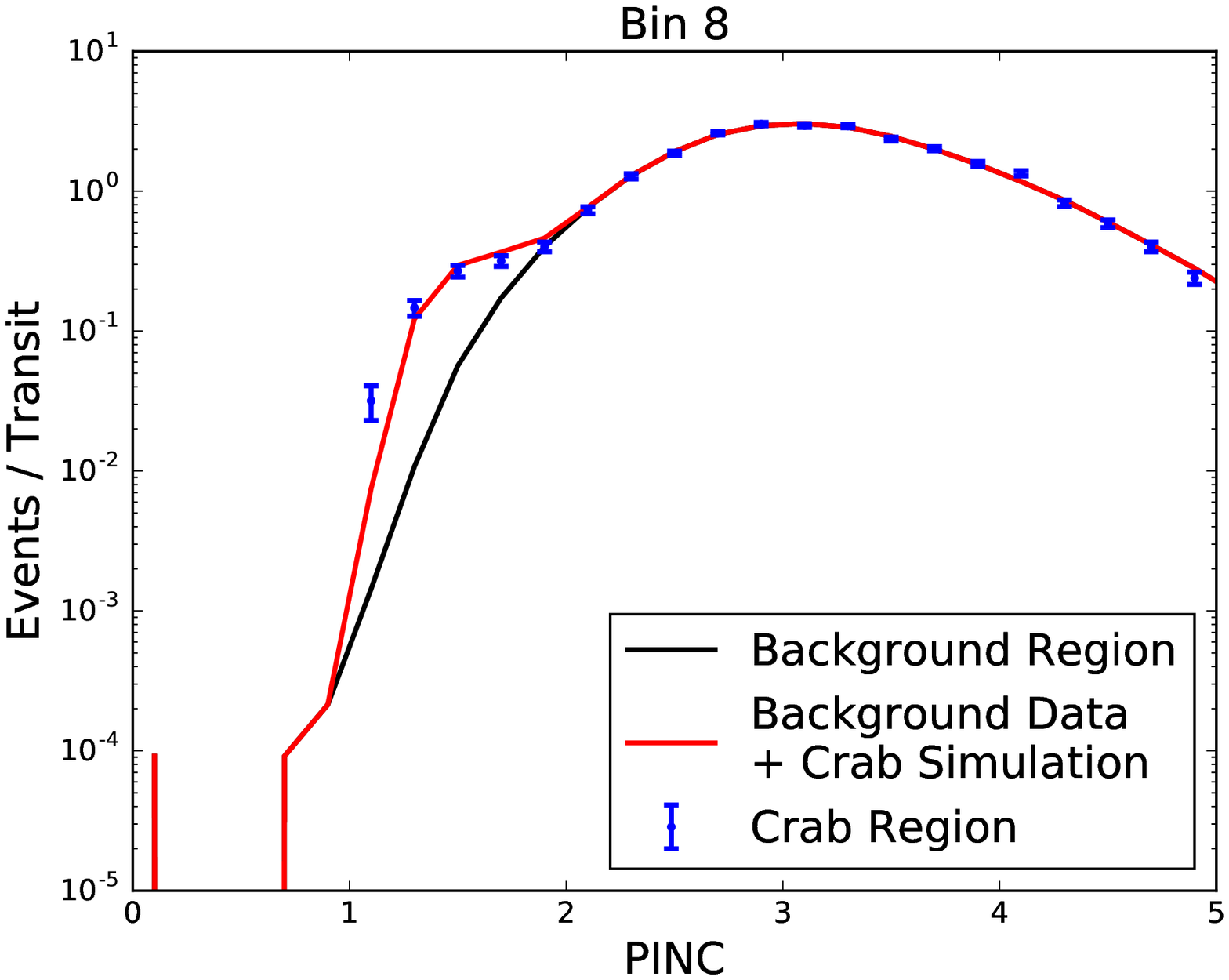}{0.45\textwidth}{(b) $\mathcal{B}=8$ Signal and Background}
          }
\gridline{\fig{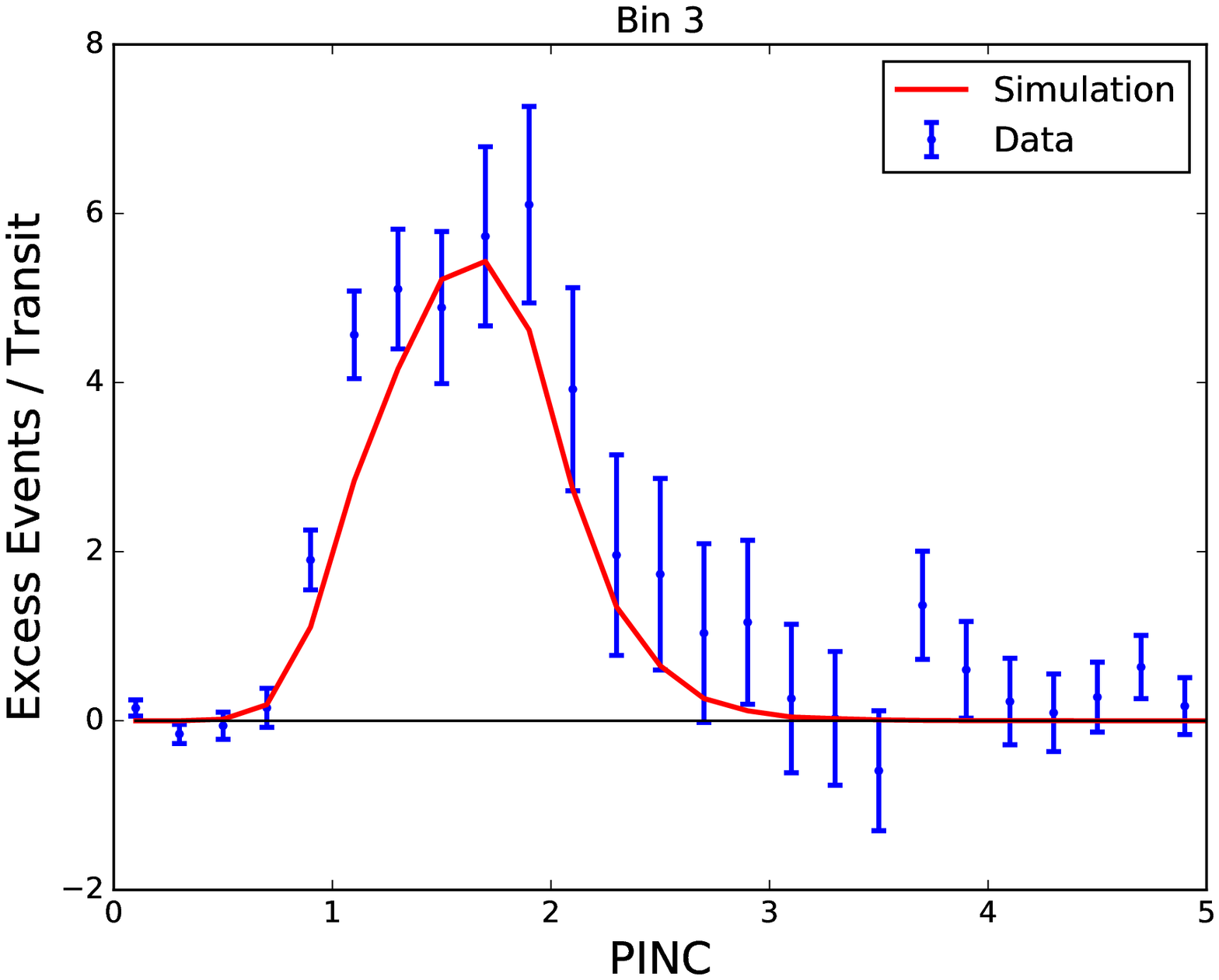}{0.45\textwidth}{(c) $\mathcal{B}=3$ Background Subtracted}
          \fig{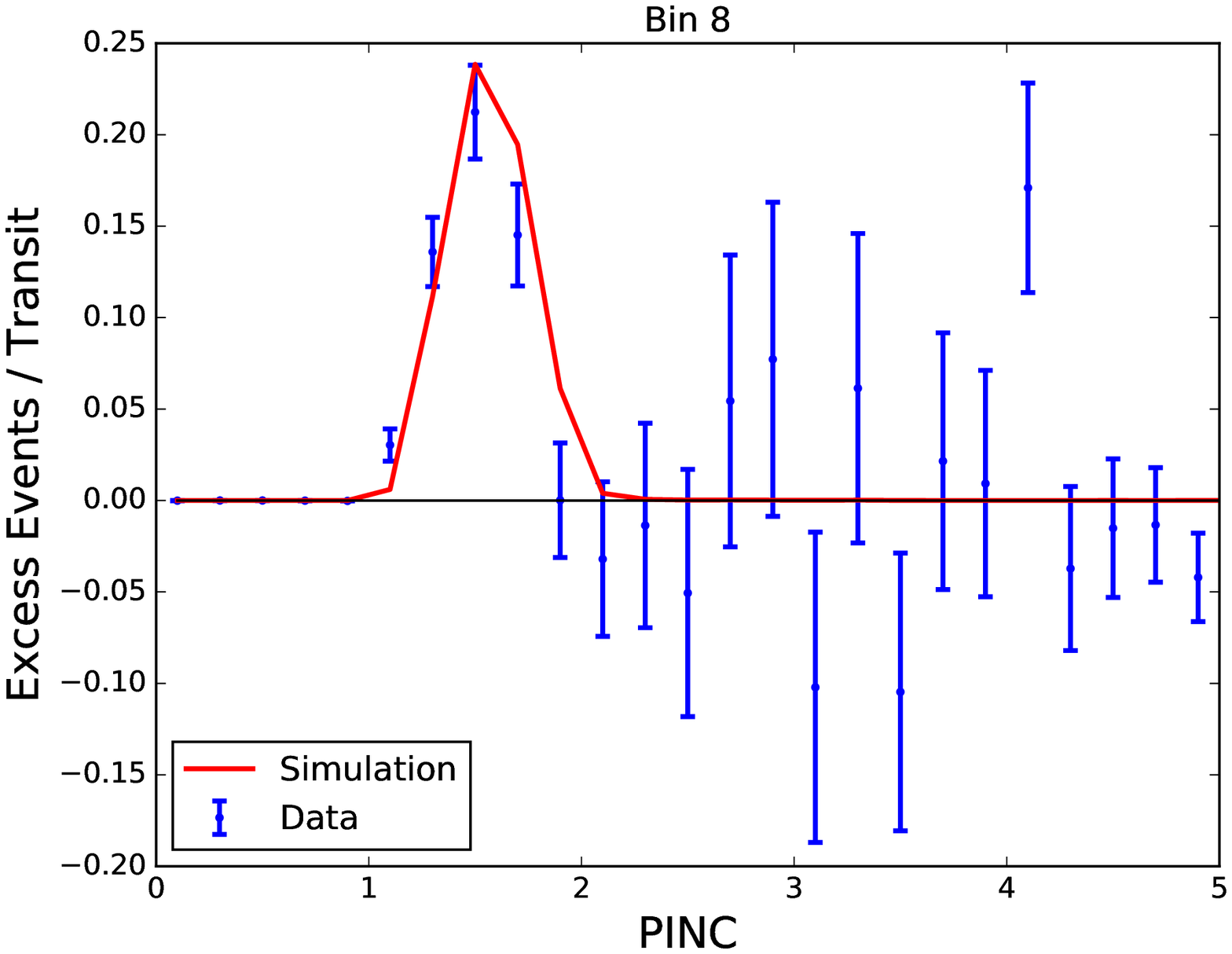}{0.45\textwidth}{(d) $\mathcal{B}=8$ Background Subtracted}
          }
\caption{Similar to Figure \ref{fig:compactdistribution}, the
PINCness, $\mathcal{P}$, distribution for events in the vicinity of the Crab
for $\mathcal{B}$=3 (left panels) and $\mathcal{B}$=8 (right panels). Since 
$\mathcal{P}$ is essentially a $\chi^2$ computation, the $\mathcal{P}$ variable
is $\sim$1-2 for true photons.
}
\label{fig:pincdistribution}
\end{figure}

The $\mathcal{C}$ and $\mathcal{P}$ variables are well-modeled in
simulation. Figure \ref{fig:compactdistribution} shows the measured
distribution of $\mathcal{C}$ in the vicinity of the Crab Nebula (the
Crab region) and
in an annular reference region around the Crab (the background region).
The background region is scaled to have the same solid angle as the Crab 
region.
The distributions
in the vicinity of the Crab are made of a combination of hadronic
cosmic-rays and true photons from the Crab. Figures
\ref{fig:compactdistribution}c and \ref{fig:compactdistribution}d
show these distributions with the background distribution subtracted. 
The subtraction yields the data-measured distribution of $\mathcal{C}$
for gamma rays from the Crab. 
Figure \ref{fig:pincdistribution} is a comparable figure for $\mathcal{P}$.
Figures \ref{fig:compactdistribution} and
\ref{fig:pincdistribution} are compared to a simulation prediction 
from the final fitted flux from Section \ref{sec:spectralfit}; 
the simulation agreement
is evident.

\subsection{Noise and Fit Refinement}
\label{sec:refinement}

HAWC's outer 8-inch PMTs individually trigger at some 20--30 kHz and the
10-inch central PMTs at 40-50 kHz. Of this random noise, roughly half (with
large uncertainties) are 
believed to be due to real shower fragments and roughly half
due to non-shower sources like radioactive contaminants in the PMT glass or
PMT afterpulsing. Approximately 1 kHz of the noise is ``high PE'' noise
from individual accidental air shower muons (10--200 PEs) and can be correlated 
between the PMTs within
a WCD that the muon hits. 

This noise can bias the air shower reconstruction and the muon noise has the
potential, if not removed, to confuse the identification of gamma-ray 
showers because a single muon is enough to indicate that a shower is of
hadronic origin. 

In order to achieve the best direction reconstruction and to avoid 
falsely rejecting true photons, 
the SFCF core reconstruction and the plane fit are each performed twice
as outlined in Table \ref{tab:recosteps}.
During the first pass, all selected hits
are used to locate the core
and initial direction. After this first ``rough'' fit, hits that are
more than $\pm$50 ns from the curvature/sampling corrected air shower 
plane are removed and the shower is fit a second time. 

The computation of photon/hadron 
separation variables 
is done with hits that are within $\pm$ 20ns from the
curvature/sampling-corrected 
air shower plane. With these cuts, only some $\sim$4\% of gamma ray
events will have an accidental muon contributing to the photon/hadron
variable computation and risk being falsely rejected.

\section{Crab Nebula Signal}
\label{sec:crabsignal}

Once individual events are reconstructed, the identification and 
characterization of sources proceeds. 
Section \ref{sec:dataset} discusses the first 553 days of data-taking.
Section \ref{sec:bkg} describes how the residual cosmic-ray background,
after gamma/hadron separation cuts, is 
estimated in the vicinity of the Crab. Section \ref{sec:angres} describes
the validation of the angular resolution. Section \ref{sec:selection}
describes the optimization of photon/hadron discrimination cuts. 

\subsection{Dataset}
\label{sec:dataset}

\begin{figure}
\centering
\includegraphics[width=0.45\textwidth]{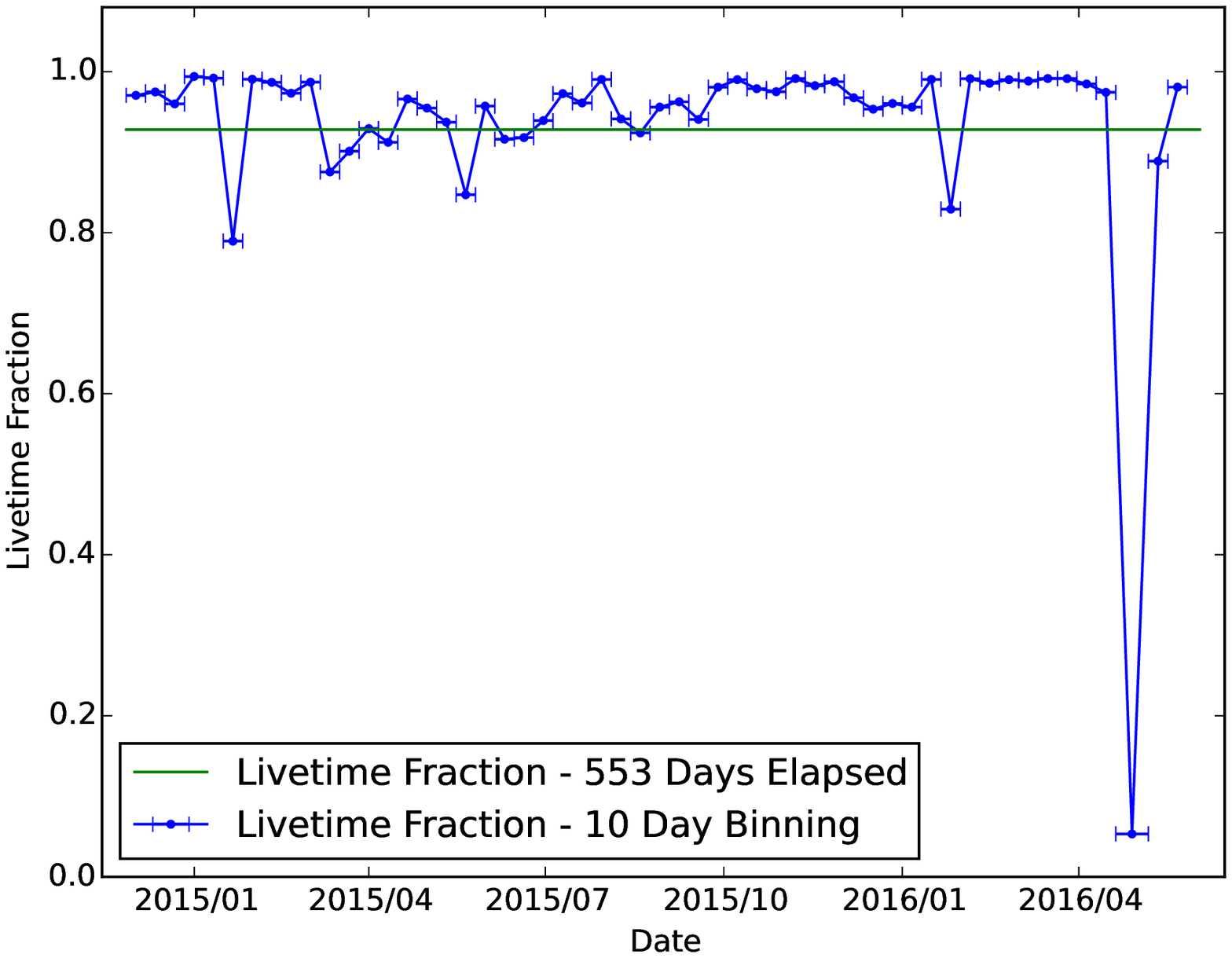}
\includegraphics[width=0.45\textwidth]{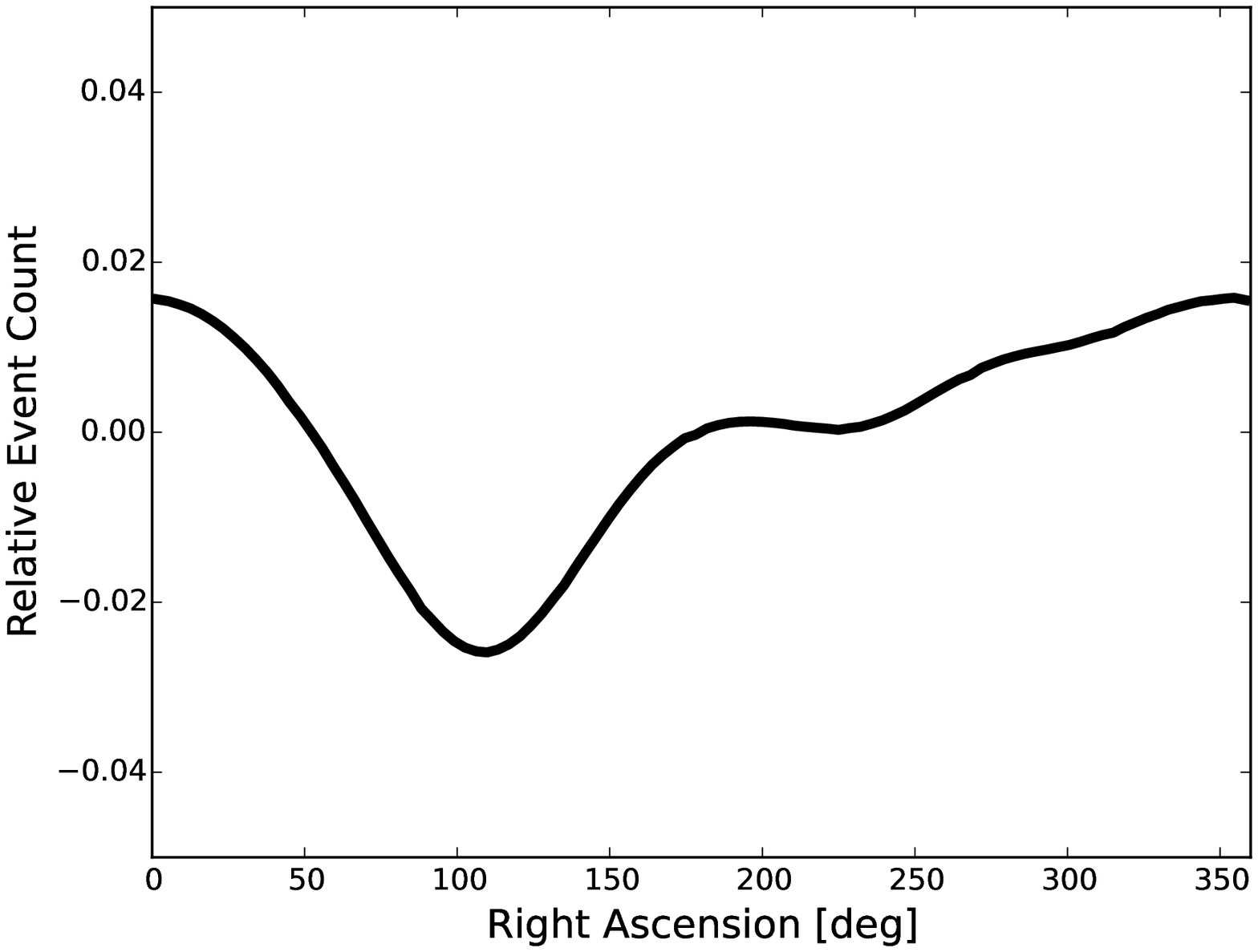}
\caption{The left panel shows the livetime fraction of HAWC 
during the first 553 days of data-taking. Data is shown averaged over 10-day 
increments along with the average from the entire dataset. 
The right panel shows the 
exposure of the instrument, measured with the reconstructed event
direction, as a function of right ascension. 
The overhead sky is 
nearly uniformly exposed with a maximum deviation from uniformity
of less than $\pm$2\%.}
\label{fig:livetime}
\end{figure}

We consider here data taken by HAWC between 2014 November 26 and 
2016 June 2, a total elapsed time of 553 days. 
The detector was not taking data for a cumulative time of 40 days for
various operational reasons. Additionally, a further 7 days of data
was rejected due to trigger rate instability.
This yields a total livetime of 
506.7 days, an average
fractional livetime of 92\%. 
Figure \ref{fig:livetime} shows the fractional livetime 
achieved in blocks of 10 days. With the exception of one period of
extended downtime (due to a failure of the 
power transformer at the site), during no single 10-day block was
the detector live less than 75\% of the time. 

The occasional downtime does not heavily bias the exposure. Figure
\ref {fig:livetime} shows the exposure of the instrument, measured
as the fractional deviation of the number of shower events observed 
as a function
of reconstructed right ascension. 
The anisotropy in cosmic-ray arrival direction is subdominant at
$\sim$10$^{-3}$ \citep{milagrolsa} and
we can 
conclude that the exposure is flat to within $\pm$2\%. 
During the 507-day livetime, the Crab is visible above a 
zenith angle of 20$^\circ$ for $\sim$1400 hours and
above a zenith angle of 45$^\circ$ for $\sim$3200 hours. 

\subsection{Background Estimation}
\label{sec:bkg}

Even with strict photon/hadron
discrimination cuts, the data is still dominated by 
hadronic cosmic-ray events. Fortunately, 
the directions of hadronic cosmic rays are randomized by
magnetic deflection in 
transit from their sources and the population of cosmic rays is isotropic
to a few parts in 10$^{3}$ \citep{milagrolsa}. Gamma-ray sources, by comparison,
appear as localized ``bumps'' on this smooth background. In order to identify
gamma-ray sources, this background contamination must be estimated. 

We utilize an algorithm developed for analysis of Milagro 
data \citep{milagrocrabspectrum}
called direct integration.
Due to the strong cosmic-ray rejection capability of HAWC, the 
number of background events observed in the highest-energy $\mathcal{B}$
bins is very sparse. To compensate for the sparseness of the background,
the direct integration background computation has been averaged
by $0.5^\circ$. This averaging has been studied using simulated
data sets and does not adversely affect the background estimate.

\begin{figure}
\gridline{\fig{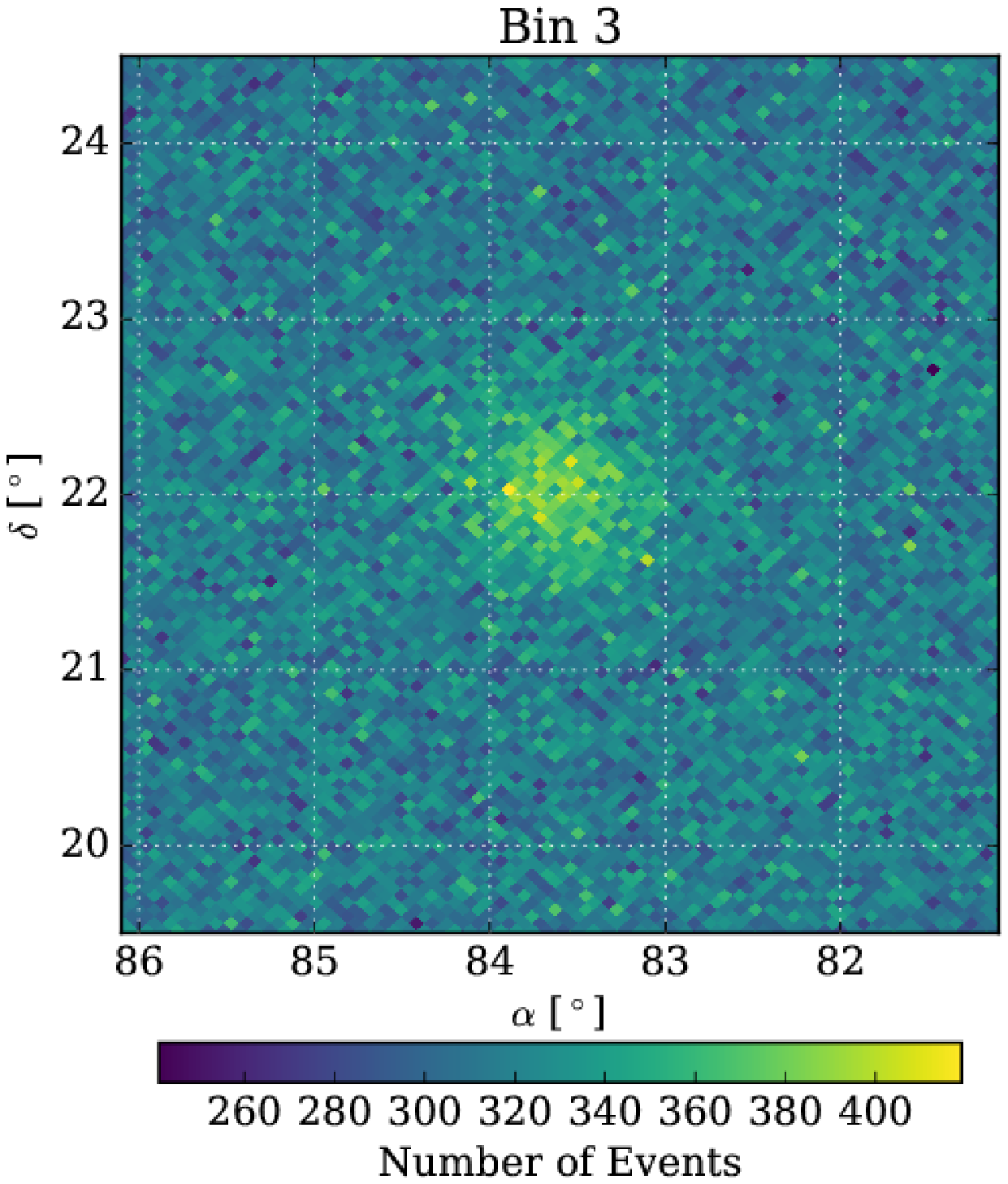}{0.45\textwidth}{(a) $\mathcal{B}=3$ Event Counts}
          \fig{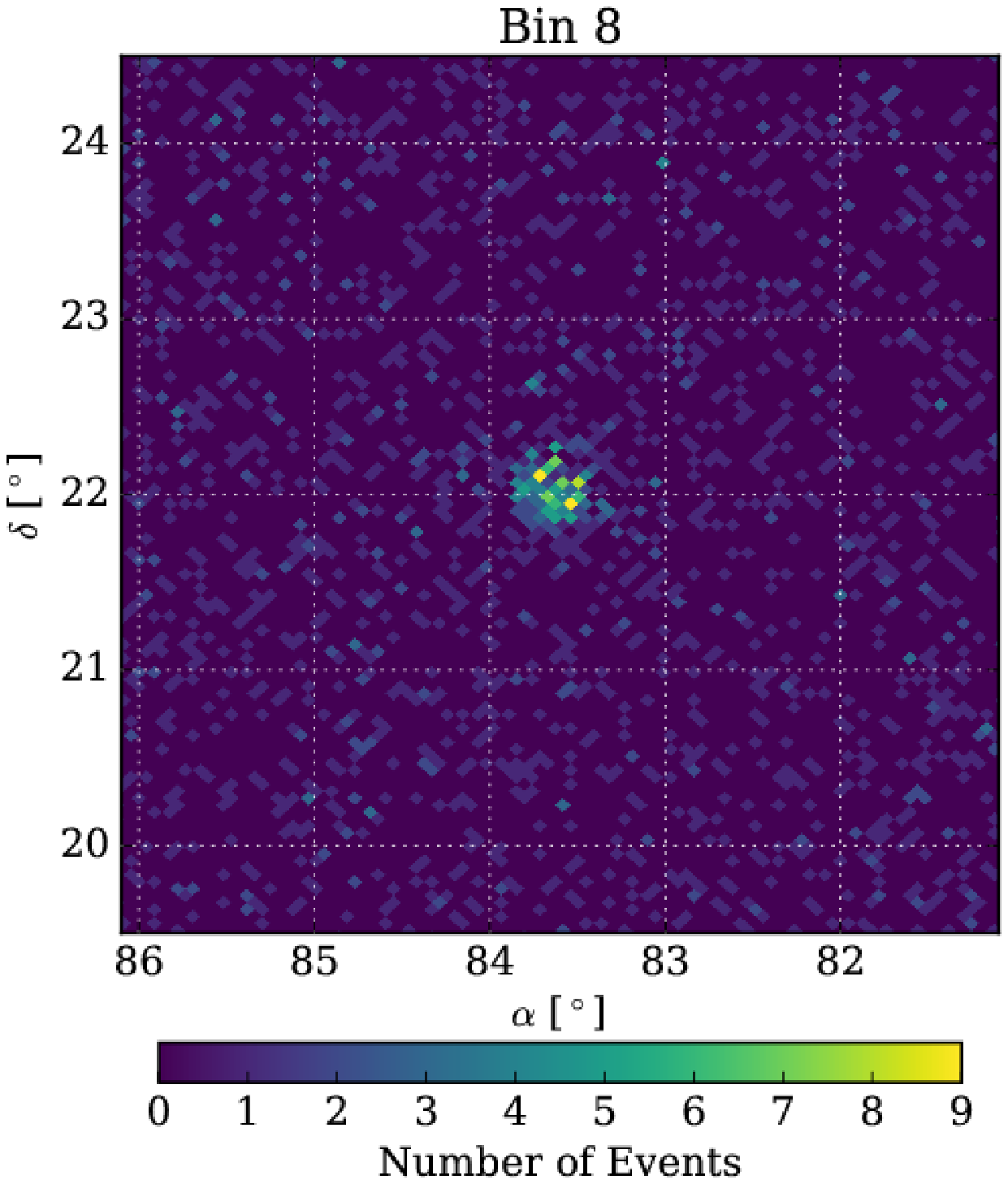}{0.45\textwidth}{(b) $\mathcal{B}=8$ Event Counts}
          }
\gridline{\fig{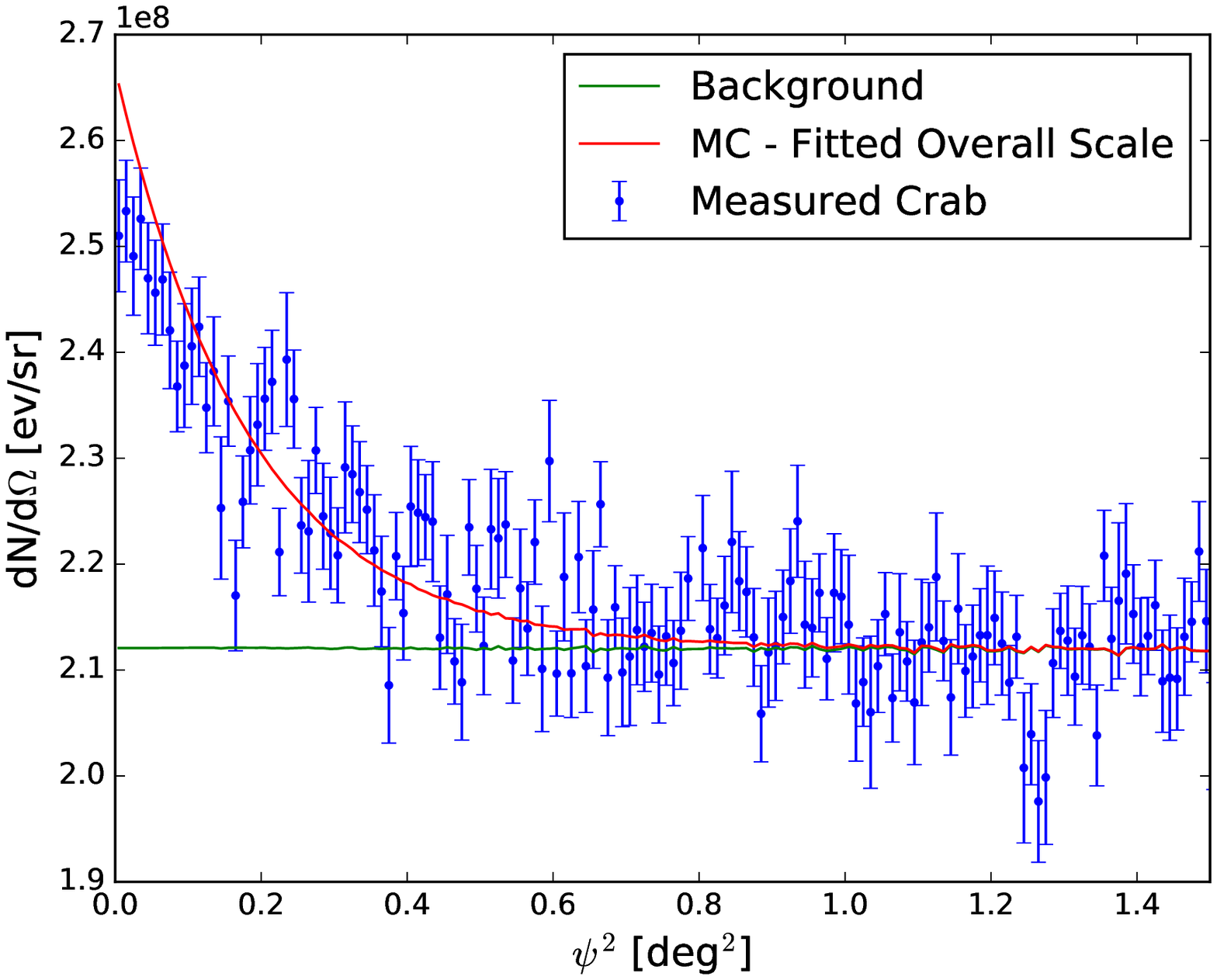}{0.45\textwidth}{(c) $\mathcal{B}=3$ Angular Profile}
          \fig{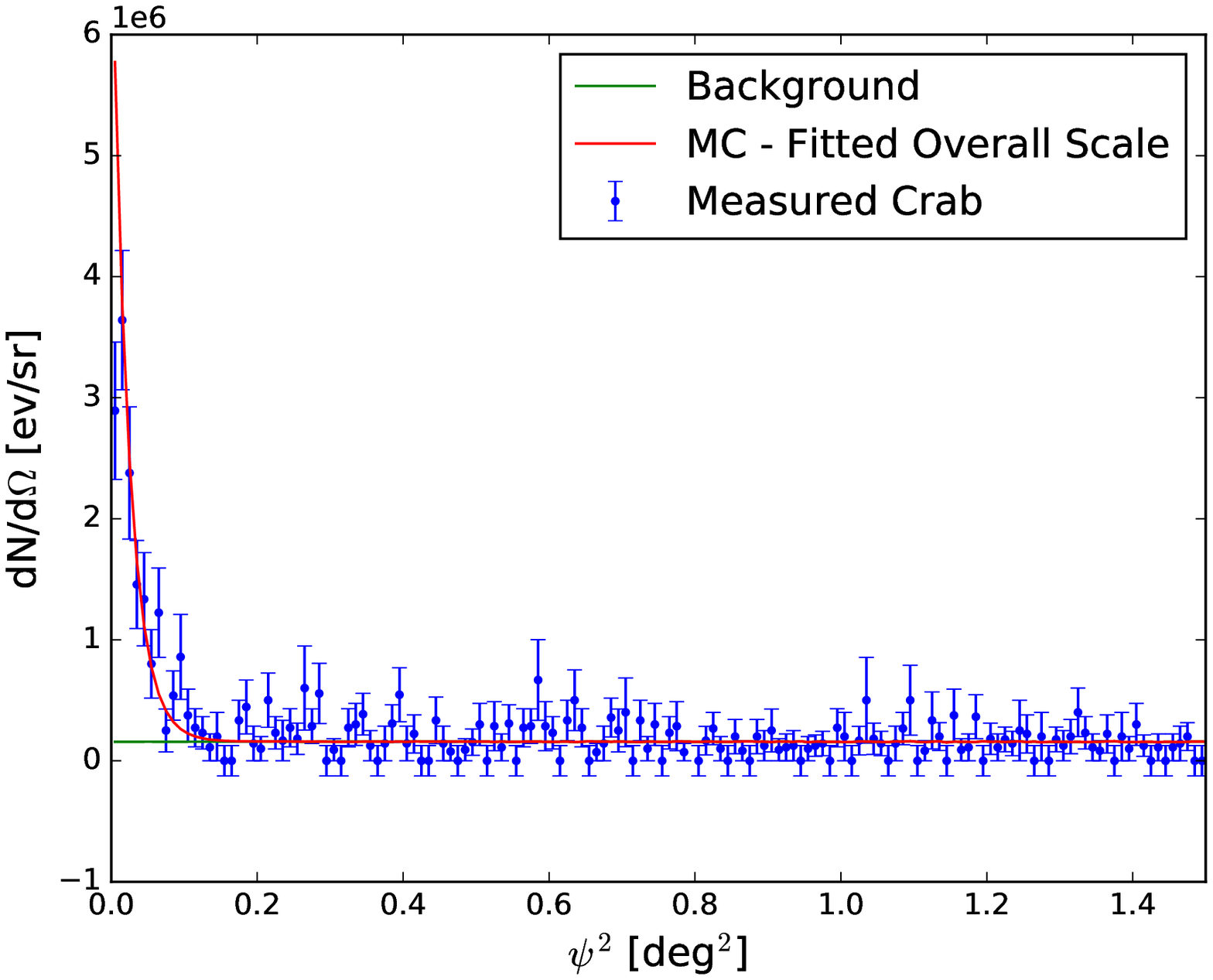}{0.45\textwidth}{(d) $\mathcal{B}=8$ Angular Profile}
          }
\caption{
Maps of the sky around the Crab Nebula for events
$\mathcal{B}$=3 (left) and $\mathcal{B}$=8 (right) after
photon/hadron discrimination in equatorial coordinates.
The top panels
show the recorded number of events pixels on the sky much
smaller than the HAWC angular resolution. The Crab is readily evident. 
The bottom panels show the number of recorded
events per steradian (dN/d$\Omega$)
as a function of the distance from the Crab. At higher $\mathcal{B}$,
the angular resolution and background rejection improve dramatically.
}
\label{fig:angresexhibit}
\end{figure}

Knowing the background at each pixel in the sky, we can evaluate the likelihood
that there is a gamma-ray source at a specific location and the 
photon flux from each source.

\subsection{Point Spread Function}
\label {sec:angres}

The point spread function, $\rho(\psi)$, describes how accurately the 
directions
of gamma-ray events are reconstructed. Here, $\psi$ is the 
space-angle difference between the true photon arrival direction and the 
reconstructed direction. To a good approximation, 
the point spread function of HAWC is the sum of two 2-dimensional Gaussians with
different widths.

\begin{equation}
\rho(\psi) = \alpha G_{1}(\psi) + (1-\alpha) G_{2}(\psi)
\label{fcn:psf}
\end{equation}

\noindent where $G_{i}$ is a Gaussian distribution with width $\sigma_{i}$

\begin{equation}
G_{i} (\psi)= {1 \over{ {2 \pi \sigma_{i}^2}}} {e^{-{\psi^2 \over {2\sigma_{i}}^2 }}}
\end{equation}
which is normalized to unity across the unit sphere. 

Figure \ref{fig:angresexhibit} exhibits the measured angular resolution in HAWC data in two 
size bins $\mathcal{B}=3$ and $\mathcal{B}=8$.
The solid-angle density of recorded events $dN/d\Omega$ in the vicinity of the Crab is shown as a function of $\psi^2$. 
Bins of $\psi^2$ have constant solid angle (in the small-angle approximation), so any remaining cosmic-ray 
background shows up as a flat 
component and the gamma rays are evident as a peak near $\psi^2=0$. The improvement in angular resolution
for larger events is clear.

\begin{figure}[h]
\centering
\includegraphics[width=0.65\textwidth]{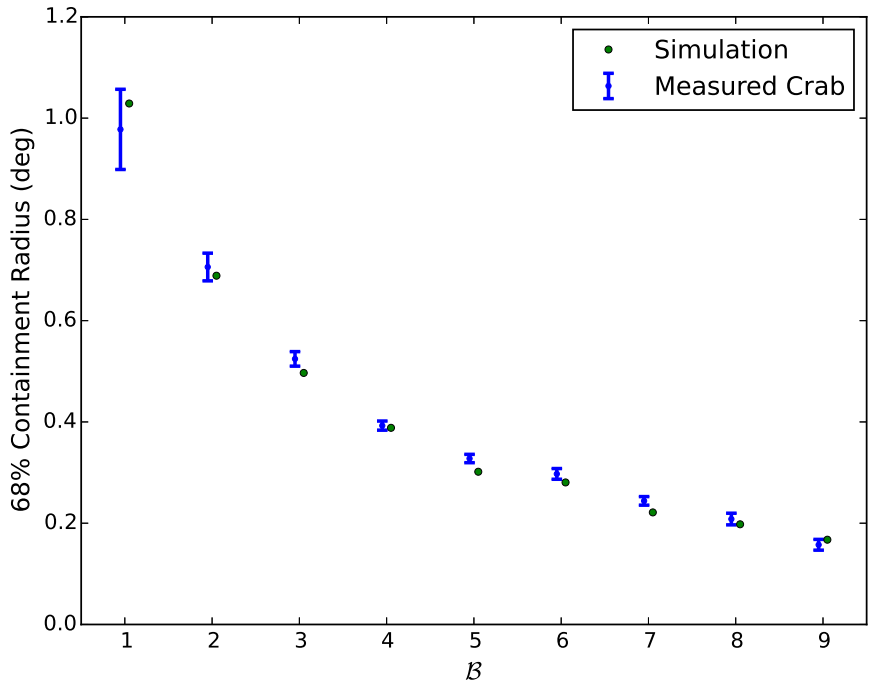}
\caption{The figure shows the measured angular resolution, the
angular bin required to contain 68\% of the photons from the Crab,
as a function of the event size, $\mathcal{B}$. 
The measurements are compared to simulation. 
The measured and predicted angular resolutions
are close enough that that using the simulated angular resolution
for measuring spectra
is a sub-dominant systematic error.  }
\label{fig:angres}
\end{figure}

Fits to this functional form of Equation \ref{fcn:psf} can have highly coupled parameters.
It is more useful and traditional
to quantify the 
resulting fits with the 68\% containment radius, $\psi_{68}$, 
the angular radius around
the true photon direction in which 68\% of events are reconstructed. 
Figure
\ref{fig:angres} shows $\psi_{68}$, for each $\mathcal{B}$ 
of the analysis, measured on the Crab and predicted from simulation.
At best, events are localized to within 0.17$^\circ$, the best angular
resolution achieved for a wide-field ground array.

Knowing the angular resolution is critical to subsequent steps of the 
analysis.
Figure \ref{fig:angresexhibit} 
indicates that the simulated angular resolution is in good 
agreement with measurements of the Crab Nebula. This is important 
because the angular resolution of HAWC for objects at declinations 
above and below the Crab will differ. While the measured PSF at 
the position of the Crab cannot be easily extrapolated to other 
declinations, the simulation can be used to predict the shape of the 
PSF at any declination. Therefore, the data-simulation agreement 
shown in Figure \ref{fig:angresexhibit} is an important verification step.

\subsection{Cut Selection and Gamma-Ray Efficiency}
\label{sec:selection}

The two parameters described in Section \ref{sec:ghsep}, the compactness, 
$\mathcal{C}$, and PINCness, $\mathcal{P}$,
are used to remove hadrons and keep gamma rays. Events are removed using 
simple cuts on these
variables and the cuts depend on the size bin, $\mathcal{B}$, of the event. 
The cuts are chosen to maximize the statistical significance with which
the Crab is detected in the first 337 days of the 507-day dataset.
Concerns of using the data itself for optimizing 
the cuts are minimal with a source as significant as the Crab. 

\begin{figure}[h]
\centering
\includegraphics[width=0.65\textwidth]{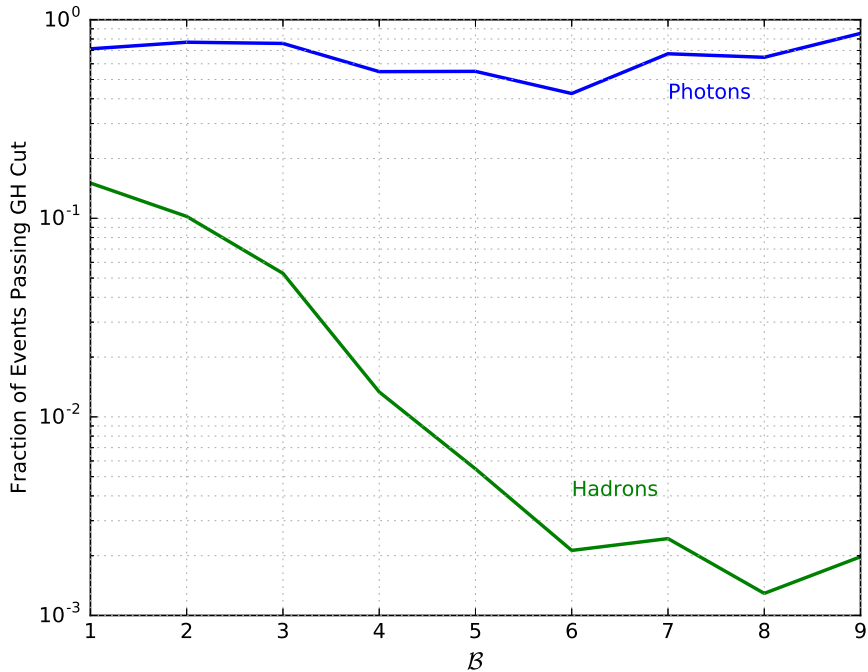}
\caption{The figure shows the fraction of gamma rays and background
hadron events passing photon/hadron
discrimination cuts as a function of the
event size, $\mathcal{B}$. Good efficiency for photons is maintained
across all event sizes with hadron efficiency approaching
1$\times$10$^{-3}$ for high-energy events. }
\label{fig:ghsep}
\end{figure}

Table \ref{table:cuts} shows the cuts chosen for each $\mathcal{B}$
bin. 
The rates of events
across the entire sky going into the 9 bins, after hadron rejection
cuts, vary dramatically,
from$~$$\sim$500 Hz for $\mathcal{B}$=1 to $\sim$0.05$~$Hz for $\mathcal{B}$=9.
Figure \ref{fig:ghsep}
shows the predicted efficiency for gamma rays (from simulation) along
with the measured efficiency for hadronic background under these cuts. The
efficiency of photons is universally greater than 30\% while keeping,
at best, only 2 in $10^3$ hadrons. 
The efficacy of the cuts is a strong function of the event size, 
primarily because
larger cosmic-ray events produce many more muons than 
gamma-ray events of a similar size. 

The limiting rejection at high energies is 
better than predicted 
in the sensitivity design study \citep{sensipaper}. The original 
study was conservative in estimating the rejection power that 
HAWC would ultimately achieve. With more than a year of
data, we now know the hadron rejection of the cuts and can accurately
compute the background efficiency.

\section{Spectral Fit}
\label{sec:spectralfit}

Knowing the angular resolution and the background in each $\mathcal{B}$, the
energy spectrum of the Crab Nebula may be inferred from the measured data. 
Section \ref{sec:likelihood} describes the likelihood
fit to the data. Section \ref{sec:likelihoodresults} describes the 
resulting measurement, and 
Section \ref{sec:systematics} describes the systematic errors to which
this measurement is subject.

\subsection{Likelihood Analysis}
\label{sec:likelihood}

The HAWC data is fit using the maximum likelihood approach
to find the physical flux of photons from the Crab \citep{wilkstheorem,liff}. 
In this approach, the likelihood
of observations is found under two ``nested'' hypotheses where some number 
of free parameters are fixed in one model. 
This approach can be used to conduct a likelihood ratio test by 
forming a test statistic, TS,
that indicates how likely the data is under 
a pure background hypothesis or to test the improvement of having additional
free parameters in the functional form of the hypothesis spectrum.

The likelihood function is formed over the small (on the scale 
of the angular resolution) spatial pixels within 2
degrees of the Crab.
Each pixel, $p$ has an expected number of background events of $B_{p}$ and, for a specific flux model, an
expected number of true photons $S_{p}(\vec{a})$, where $\vec{a}$ denotes the 
parameters of our spectral 
model of the Crab. The predicted photon counts fall off from the source
according the assumed point spread function. The 
likelihood  $\mathcal{L}(\vec{a})$ is then the simple Poisson probability
of obtaining the measured events in each pixel, $M_p$ under the assumption of the flux
given by $\vec{a}$. The $\mathcal{B}$ dependence of each term in 
Equation \ref{eqn:likelihood} is suppressed. 

\begin{equation}
{\rm{ln}}(\mathcal{L}(\vec{a})) = \sum_{\mathcal{B}=1}^{9} \sum_{p=1}^{N} {\rm{ln}} \left(  {{( B_{p} +S_{p}(\vec{a})  )} ^{M_p} e^{B_{p} +S_{p}(\vec{a}) } \over {M_p!}} \right)
\label{eqn:likelihood}
\end{equation}

\noindent Specifically, we fit a differential photon flux $\phi(E)$ of the log parabola (LP) form: 

\begin{equation}
\phi(E) = \phi_0 (E/E_{0})^{-\alpha -\beta\cdot{\rm{ln}}(E/E_{0})}
\label{eqn:logparabola}
\end{equation}

\noindent Here, $\phi_0$ is the flux at $E_{0}$, 
$\alpha$ is the primary spectral index and $\beta$ is a second 
spectral index that governs the changing spectral power across the energy range of the fit. 
In this formulation, $E_{0}$ is not fitted but is chosen to 
minimize correlations between the free parameters in the fit.
When fit to an LP function,
$E_{0}=7~{\rm{TeV}}$ produces good results.

\subsection{Fit Results}
\label{sec:likelihoodresults}

\begin{figure}[h]
\centering
\includegraphics[width=0.3\textwidth]{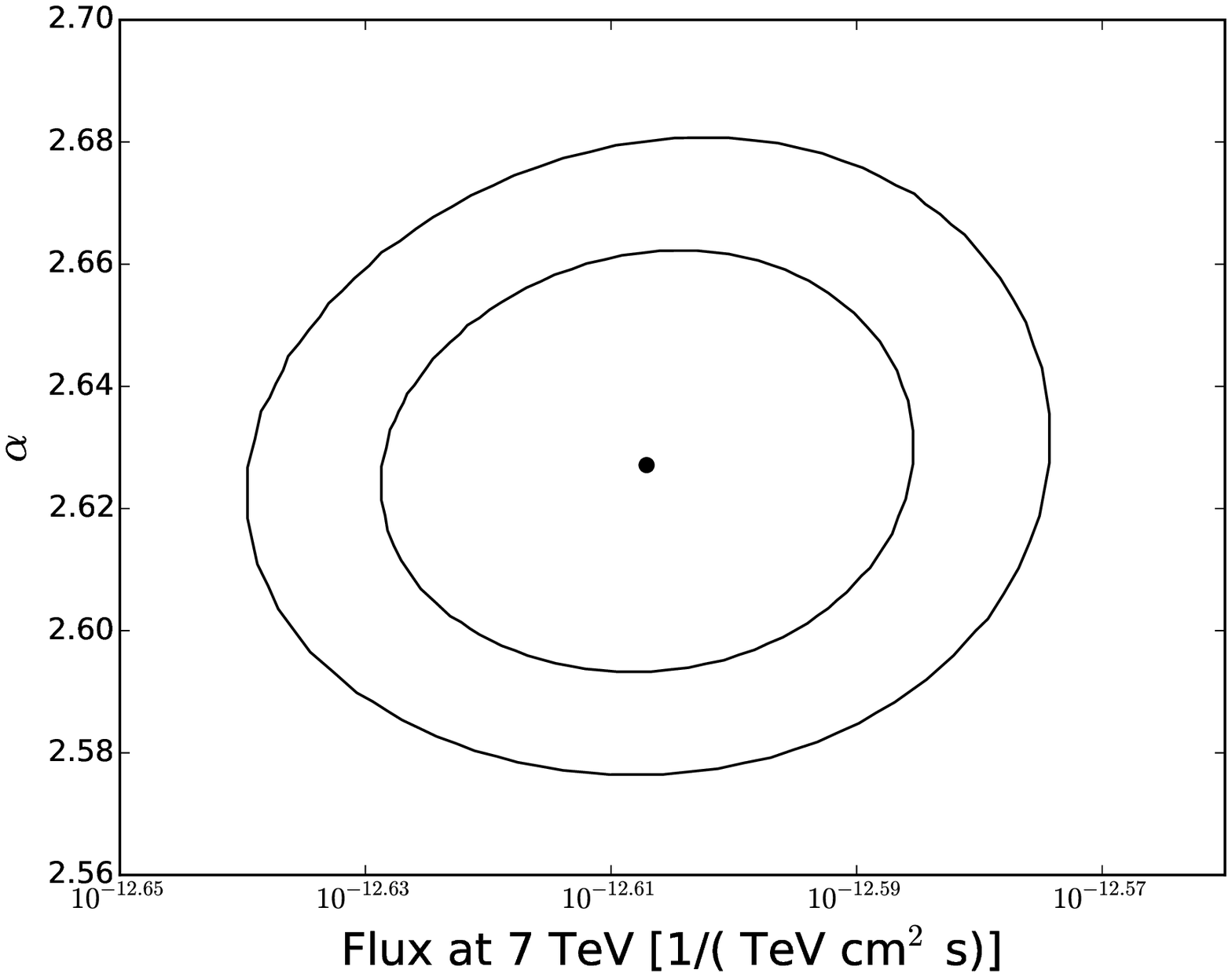}
\includegraphics[width=0.3\textwidth]{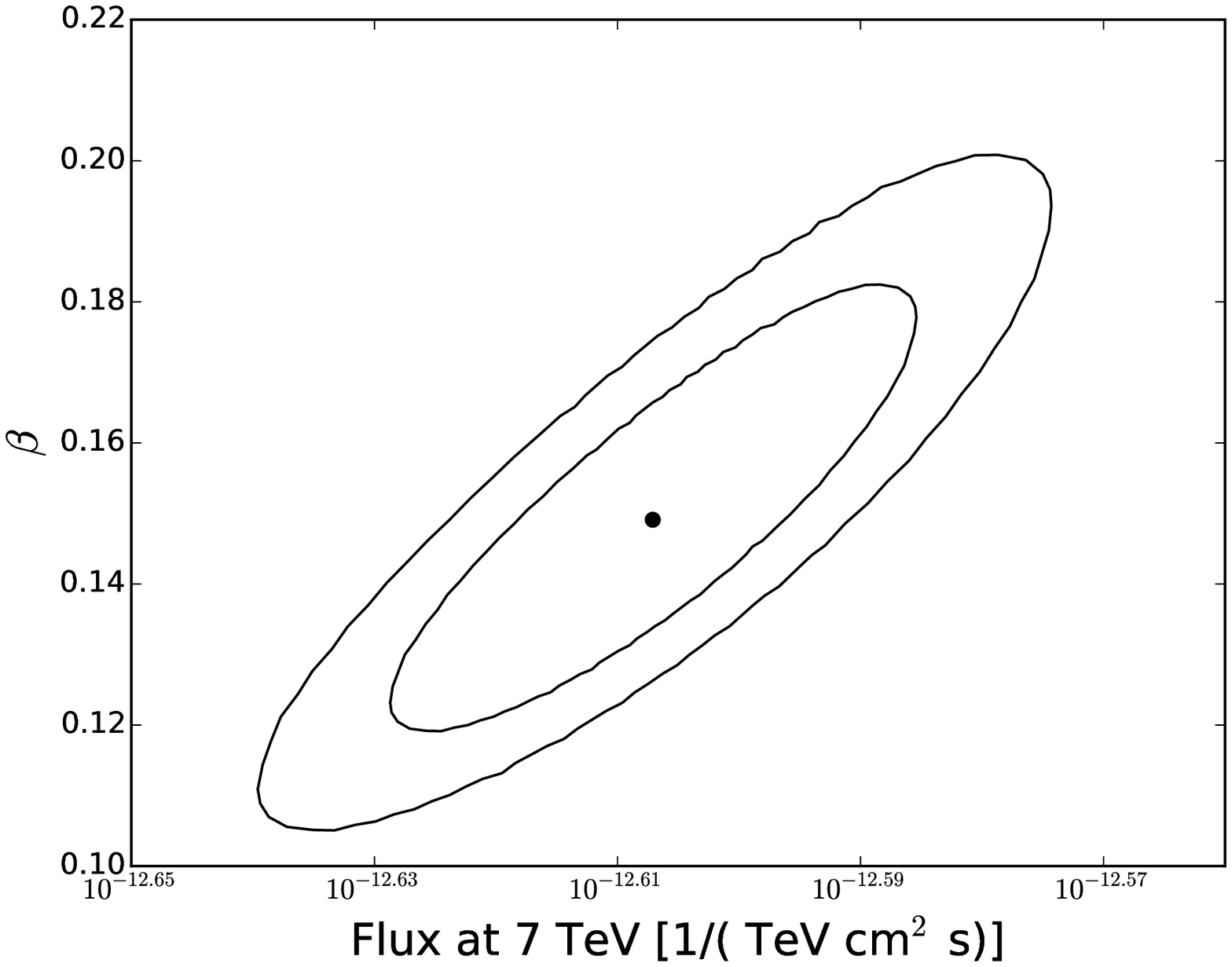}
\includegraphics[width=0.3\textwidth]{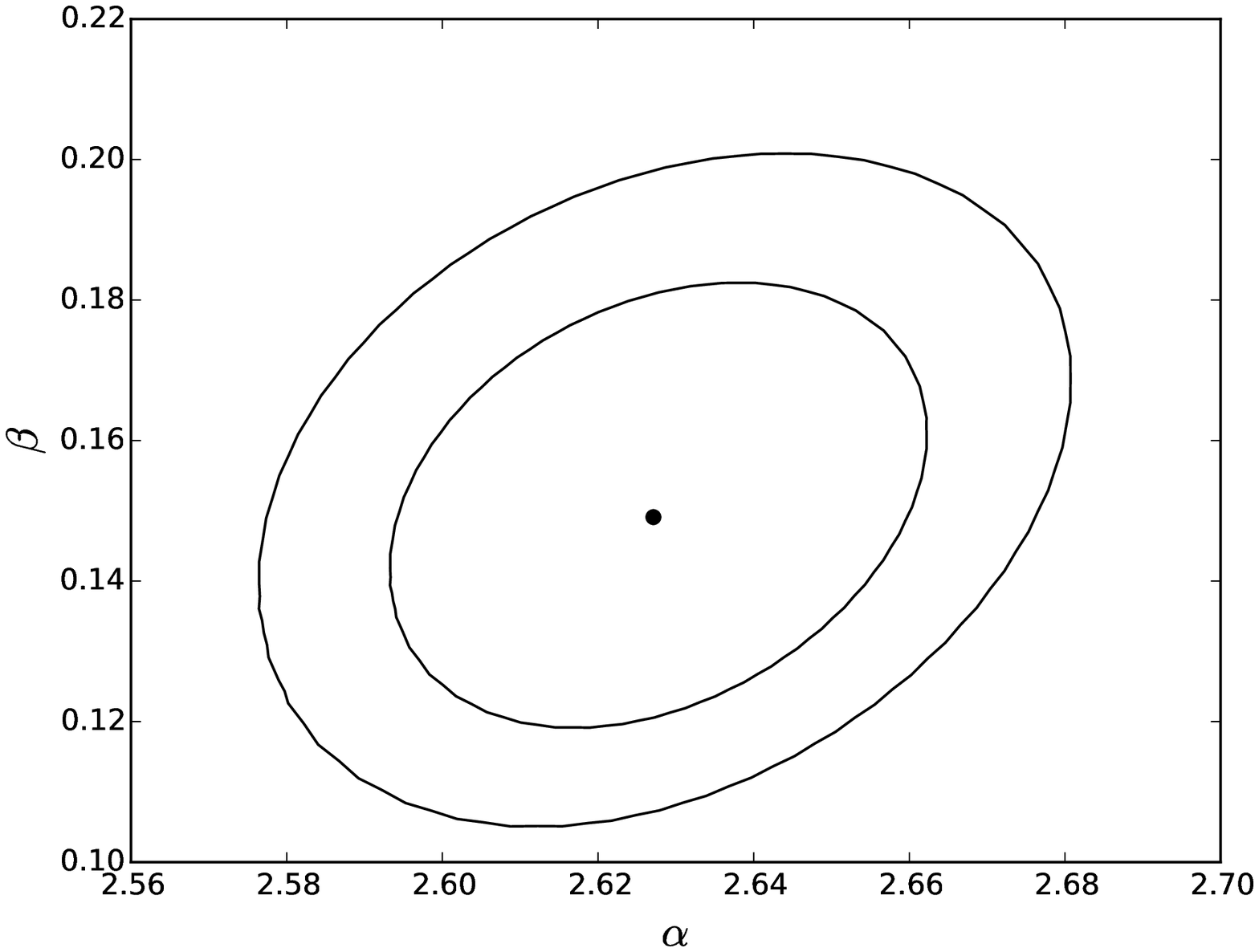}
\caption{Likelihood space around Crab best fit. Shown are the best-fit
flux and the region of fluxes 
allowed at 1 and 2$\sigma$. The space shown is the 
$\phi_0$, $\alpha$ and $\beta$ space from Equation \ref{eqn:logparabola}
with a pivot energy of $E_{0}=$7 TeV. 
}
\label{fig:likelihoodspace_logparabola}
\end{figure}

We find the parameters for $\vec{a}$ that maximize the likelihood function under 
signal and background hypotheses
and quantify the error region of $\vec{a}$ using Wilks' Theorem \citep{wilkstheorem}.
Figure \ref{fig:likelihoodspace_logparabola} shows the corresponding spaces of 
$\alpha$, $\beta$ and $\phi_0$ for the LP fit
that are consistent with HAWC data at 1 and 2$\sigma$.
The maximum 
likelihood occurs at $\alpha=2.63\pm0.03$, $\beta=0.15\pm0.03$, and 
log$_{10}(\phi_0~{\rm{cm}^2}~{\rm{s}}~{\rm{TeV}})=-12.60\pm0.02$.
At this best flux the TS, compared to the background-only hypothesis, is 11225, a more than 100$\sigma$ detection. 

The TS between
an unbroken power-law hypothesis (with $\beta=0$) and the full LP fit is 
142, so the spectrum is inconsistent with
an unbroken power law at 12$\sigma$.

We quantify the energy range of this fit two ways. First, we take 
the spectral fit solution and 
compute the 
lower 10\% quantile of true energy for $\mathcal{B}$=1 and
the upper 90\% quantile of true energy for $\mathcal{B}$=9. These
are 375 GeV and 85 TeV respectively. This is the energy range over which,
under the fitted hypothesis, most of the measured photons are expected
to lie.

A more conservative approach can be made focusing on the lowest and highest
energies where HAWC data could definitively reveal a sharp 
cutoff in the spectrum. 
To do this, we
separately fit functions of the forms:

\[
    \phi(x)= 
\begin{cases}
    0 & \text{if } E\geq E_{\rm{high}} \\
    \phi_{0} E^{-\alpha},              & \text{otherwise}
\end{cases}
\]
 and 

\[
    \phi(x)= 
\begin{cases}
    0 & \text{if } E\leq E_{\rm{low}} \\
    \phi_{0} E^{-\alpha},              & \text{otherwise}
\end{cases}
\]

\noindent to find the highest $E_{\rm{low}}$ and lowest $E_{\rm{high}}$ that are, 
at 1$\sigma$, inconsistent with the HAWC observation. With this approach,
we believe that we have positive detection of photons from the Crab
between 1 and 37 TeV. This is not to say
that higher or lower energy photons cannot be a part of the HAWC 
observation, but using the event size
$\mathcal{B}$ to measure the 
energy of photons limits the dynamic range of the observation. Other sources
at other declinations may yield different answers.

\begin{figure}[h]
\centering
\includegraphics[width=0.65\textwidth]{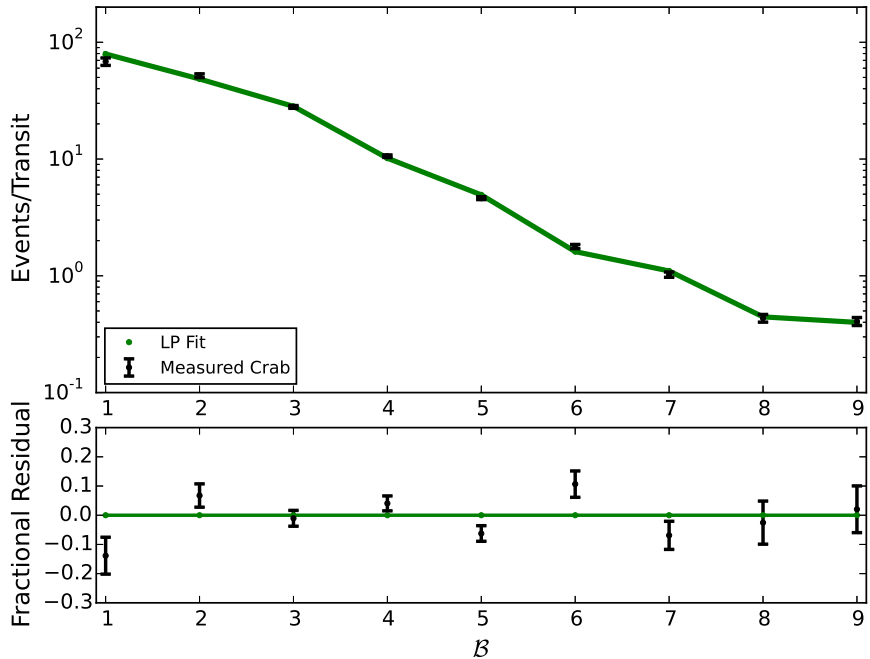}
\caption{The figure shows the measured, background-subtracted 
number of photons from the Crab
in each $\mathcal{B}$ bin. To get the total number of photons,
the signal from the Crab is fit for each $\mathcal{B}$ separately. 
The measurements are compared to prediction from simulation assuming
the Crab spectrum is at the HAWC measurement. The fitted spectrum
is a good description of the data, with no evidence of bias in the
residuals. }
\label{fig:excess}
\end{figure}

\subsection{Systematic Errors}
\label{sec:systematics}

Table \ref{tab:systematics} summarizes the major systematic errors contributing
to the measurement of the Crab spectrum with HAWC. These systematic errors have 
been investigated
by computing the spectrum from the Crab under varying assumptions to study the stability
of the results under perturbation of the assumptions.

For spectral measurements, a systematic error in three quantities is shown: 
the overall flux, the spectral index measured and 
the energy scale. The errors are summed in quadrature to 
arrive at a total systematic error.
In addition to these systematic errors, a systematic error in 
the absolute pointing of the instrument has been
studied. 

\subsubsection{Charge Resolution and Relative Quantum Efficiency}

The charge resolution is a quantity that captures by how much individual PMT charge measurements
can vary, for fixed input light, and is estimated to be 10-15\% from 
studies 
using the HAWC calibration system. Additionally, PMTs vary in their 
photon detection efficiency by
15--20\%. These factors are not simple numbers, but vary
for different light levels in the detector and can 
change with the arrival time distribution during air showers.
Varying these assumptions, the $\mathcal{P}$ and $\mathcal{C}$ change and 
impact the event passing rates, impacting the spectrum. 

\subsubsection{PMT Absolute Quantum Efficiency}

PMTs have an efficiency for converting photons impinging on their surface into PEs
detected by the PMT, typically between 20-30\%. Of course a single 
``efficiency'' number vastly simplifies
the situation: the efficiency is divided between the efficiency for producing a PE
and the efficiency for collecting a PE, varies across the face of the PMT, and is wavelength dependent. Additionally,
the absorption of the water itself is wavelength dependent. 
Much of this is modeled, but the simulation carries uncertainties in the treatment and is difficult
to validate. The calibration system, in particular, cannot yield the absolute PMT efficiency because
it requires establishing the efficiency of the calibration system's 
optical path to the PMTs much more precisely than is known. Furthermore, the 
laser for the calibration system is green light and must be extrapolated for 
application to blue Cherenkov light. 

Instead, the absolute efficiency is established by 
selecting vertical muons in HAWC tanks by their timing properties.
Vertical muons are typically minimum ionizing with a 
relatively constant energy loss.
The simulated response to vertical 
muons is scaled to match data. Nevertheless, we estimate a $\pm10\%$ uncertainty 
in the absolute PMT efficiency which propagates to the errors in the 
spectra of sources.

\subsubsection{Time Dependence, PMT Layout and Crab Optimization}

The HAWC instrument has changed over time. The main change is that
PMTs or channels are occasionally removed during 
maintenance.
Repaired channels have been re-calibrated and the calibration
constants 
have been changed occasionally for other reasons. 
The event size bins, $\mathcal{B}$, are based on fractions
of available PMTs to mitigate the impact of the varying numbers 
of PMTs. Furthermore, a single simulation
with a single representative PMT layout is used to model 
the detector and this simplification results in 
a corresponding systematic error.

Different detector layouts were simulated to bound the impact of sporadic
tubes being added or removed. As confirmation, the passing rate of 
background cosmic rays through the photon/hadron discrimination 
cuts was studied and shows comparable drifts to the simulation studies. 

Additionally, the cuts used in the analysis were established by maximizing 
the statistical significance of the 
observations of the Crab Nebula during the
first 337 days of data. While strictly not an {\it{a priori}} measurement, 
the Crab is strong enough in HAWC data that there is very little
bias to the final measurement from this optimization process 
(and no bias for 
other weaker sources). 

To investigate any potential bias, 
the data was divided in two pieces, a 337-day and a 170-day
dataset. The 337-day dataset corresponds to the period over which the
cut optimization was done and is the only dataset that could
have an over-optimization bias. The Crab spectrum was then measured separately in each
dataset.
The fitted spectra differ by $\pm$10\% in the flux and $\pm$0.1 in the spectral
index, similar to what is expected from the varying number of PMTs.

It is unclear whether the different Crab spectra in the 337-day and 
170-day datasets are due to overtuning the Crab or the changing detector
later in the data-taking. They are similar size effects. Whatever the origin
of the effect, it is a sub-dominant, but non-negligible, systematic error.

\subsubsection{Angular Resolution}

The chief uncertainties in the angular resolution arise from
a mismatch between the data and the simulation and spectral dependence
of the angular resolution. The impact of angular resolution
has been studied by reconstructing the Crab spectrum under different 
angular resolution hypotheses.

\subsubsection{Late Light Simulation}

The single largest source of systematic error is how late light in the air shower is treated. Simulation suggests that the arrival time distribution 
of PEs at the
PMTs should be well within $\sim$10 ns. Nevertheless, the distributions
of $\mathcal{C}$ and $\mathcal{P}$ in background cosmic rays,
as well as the raw PE distributions themselves, suggest some mis-modeled
effect above about 50 PEs. 

Dedicated studies of late light, using the the calibration system, 
have been seen to extend ToT measurements,
thereby distorting the measured charge in PMTs, but 
an arrival time distribution much wider than expected from simulation
is needed to explain the data.

Efforts to understand
this systematic are aimed at measuring the entire PMT waveform
for a sample of PMTs to better
understand the arrival time distribution of PEs without requiring simulation. 
It is likely that this systematic error will be better understood in the future,
but currently it 
dominates. 

\begin{table}
\begin{tabular}{ p{6cm} c c c }
Systematic &  Overall Flux & Spectral Index & log$_{10}$(E) \\
\hline
Charge Resolution/ Relative Quantum Efficiency &  $\pm$ 20\% & $\pm$ 0.05  & $<\pm$ 0.1 \\
PMT Absolute Quantum Efficiency                &  $\pm$ 15\% & $\pm$ 0.05  & $<\pm$ 0.1 \\
Time Dependence, PMT Layout and Crab Optimization &  $\pm$ 10\% & $\pm$ 0.1  & $<\pm$ 0.1 \\
Angular Resolution                             &  $\pm$ 20\% & $\pm$ 0.1   &  \\
Late Light Simulation                          & $\pm$ 40\%  &  $\pm$ 0.15 & $<\pm$ 0.15 \\
\hline
Total Flux                                          &  $\pm$ 50\% & $\pm$ 0.2   & $<$ 0.2 \\
\end{tabular}
\caption{Summary of primary contributions to HAWC systematic error in 
measuring photon fluxes. The
different effects are described in the text. Systematics in the overall flux, the
spectral index of sources, and the energy scale are shown. The 
systematics claims are conservative and are likely to improve with more understanding and
better modeling.}
\label{tab:systematics}
\end{table}

\subsubsection{Absolute Pointing}

The absolute pointing error is estimated to be no more than $0.1^\circ$ for sources 
that transit above 45$^\circ$ in HAWC. It is estimated to be no more than  $0.3^\circ$
for higher-inclination sources.

The absolute pointing of the HAWC instrument is impacted by the timing calibration
as discussed in Section \ref{sec:calib}. Each PMT has a calibrated offset
to account for different cable lengths and other timing delays. These offsets 
are established coarsely by repeated reconstructions to force the peak of maximum
cosmic-ray density to be overhead. With this correction, the location of the 
Crab is within 0.2$^\circ$ of its true location. A final detailed alignment is performed
to put the Crab Nebula in its known location. The Crab itself must be in the correct location,
by construction. 

The absolute pointing error on other sources has been studied two ways. First, 
the 
Crab location has been fit 
using only events in bands of reconstructed zenith angle. The Crab location
drifts by no more than $0.1^\circ$ up to a zenith angle of 45$^\circ$. Above that inclination,
the Crab is weakly detected and we cannot independently demonstrate better than $0.3^\circ$ 
absolute pointing error. Furthermore, the Crab location has been reconstructed
separately using data from each of the 9 $\mathcal{B}$ bins and they agree to within $0.1^\circ$.

Finally, 
other bright known sources, the blazars Markarian 421 and Markarian 501, agree with their
known locations to within 0.1$^\circ$.

\section{Discussion}
\label{sec:discussion}

\subsection{Comparison to Other Experiments}

\begin{figure}[h]
\centering
\includegraphics[width=0.65\textwidth]{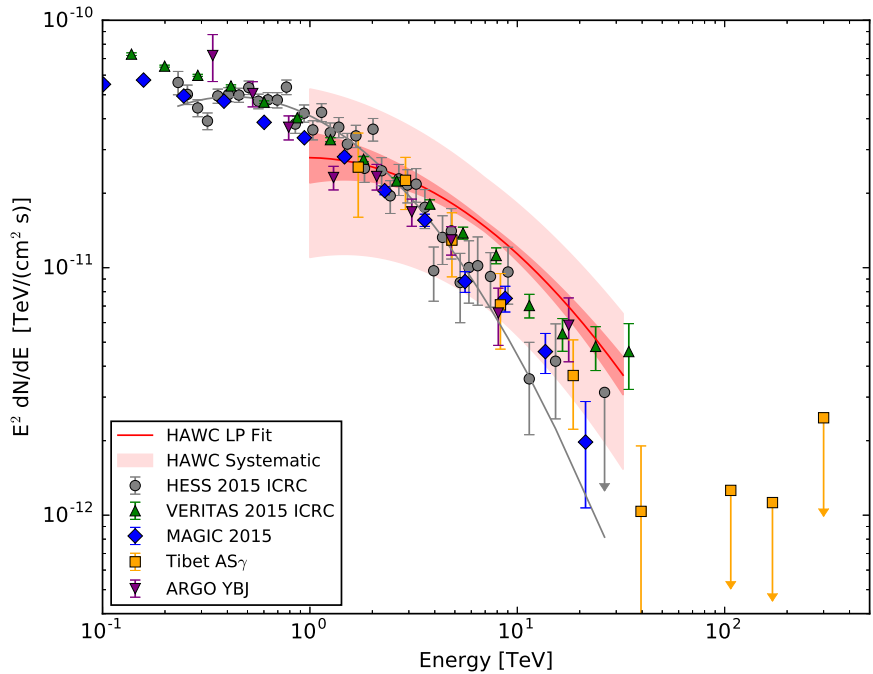}
\caption{Crab photon energy spectrum 
measured with HAWC and compared to other measurements
using other instruments \citep{hesscrab2015,magiccrabnebula,veritascrab2015,tibet100tevul,tibetcrab,argocrab} 
The red band shown for HAWC is the ensemble of fluxes allowed at 1$\sigma$
and the best fit is indicated with a dark red line. The light red band
indicates the systematic extremes of the HAWC flux. } 
\label{fig:crabfittedflux}
\end{figure}

Figure \ref{fig:crabfittedflux} shows the Crab spectrum
measured with HAWC
between 1 and 37 TeV compared to the spectrum reported by other
experiments. It is consistent with prior measurements within the systematic
errors of the HAWC measurement. 

A number of improvements to the HAWC measurement 
--- most notably the inclusion of a proper energy reconstruction --- will
reduce the systematic
errors and increase the dynamic range of our measurements. 
The observation of the Crab validates this analysis for
subsequent application to the sources across the rest of the sky. 

\subsection{Performance Figures}

\begin{figure}[h]
\centering
\includegraphics[width=0.65\textwidth]{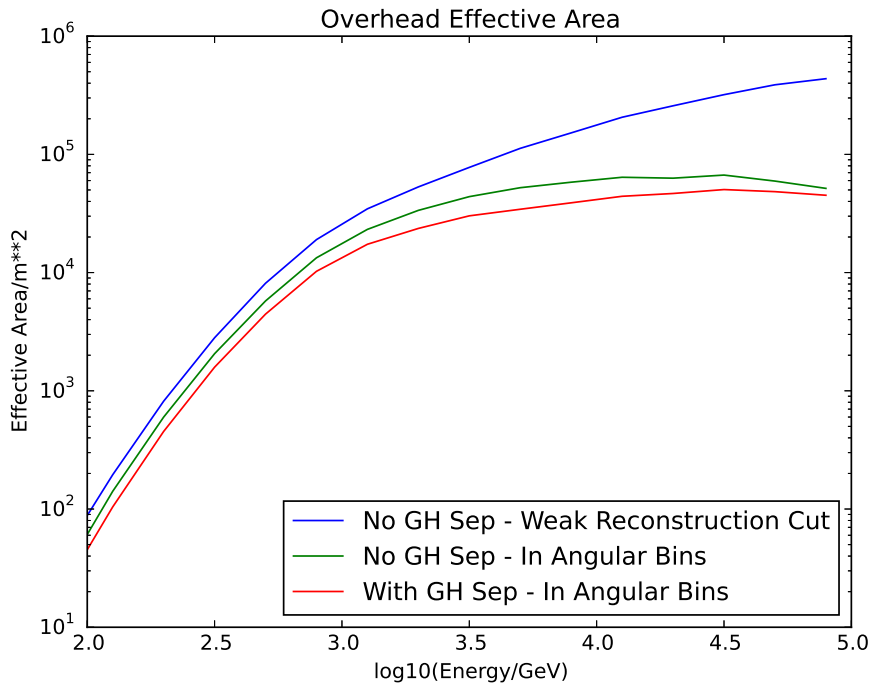}
\caption{The effective area for HAWC for events within 13$^\circ$ from overhead.
To show the progression of analysis cuts, we show curves without any 
photon/hadron discrimination, insisting that events only reconstruct
within 4$^\circ$ of their true direction. Requiring events to be 
reconstructed within their 68\% containment radius lowers the effective
area and photon/hadron discrimination cuts lowers it further. With
a requirement that events be reconstructed on the detector, the 
effective area flattens at roughly half the physical area of the instrument.}
\label{fig:aeff}
\end{figure}

Figure \ref{fig:aeff} shows the effective area of HAWC, in this analysis, 
to photons arriving within 13$^\circ$ from overhead. 
The effective area is defined as the geometrical area over which events 
are detected, convolved with the efficiency for detecting events. 
The exact conditions for a photon to be considered detected are complicated 
for this analysis because, since we are performing
a likelihood fit of all pixels in the vicinity of the Crab, 
photons that are poorly reconstructed play some role in the analysis. 
In order to have a well-formed effective area, we consider only photons 
defined within the 68\% containment radius, from Table \ref{tab:recosteps}. 
For comparison, Figure \ref{fig:aeff} includes the progression of cuts, 
from the effective area without any photon/hadron discrimination without
a strong angular accuracy cut to the full analysis cuts. The effective 
area can exceed the geometrical area of the instrument 
(about 2$\times10^4$ m$^2$)
because events with a 
core location off the detector occasionally pass the imposed cuts. 

This computation of the 
effective area is lower than in \citep{hawcgrbsensi}.
For this analysis, 
developed for steady, multi-TeV sources, we
employ tight angular cuts and have a higher energy threshold 
than used in the initial design study. Furthermore, the re-triggering,
defined by the lower edge of $\mathcal{B}=1$, limits the effective
area below 1 TeV.

\begin{figure}[h]
\centering
\includegraphics[width=0.80\textwidth]{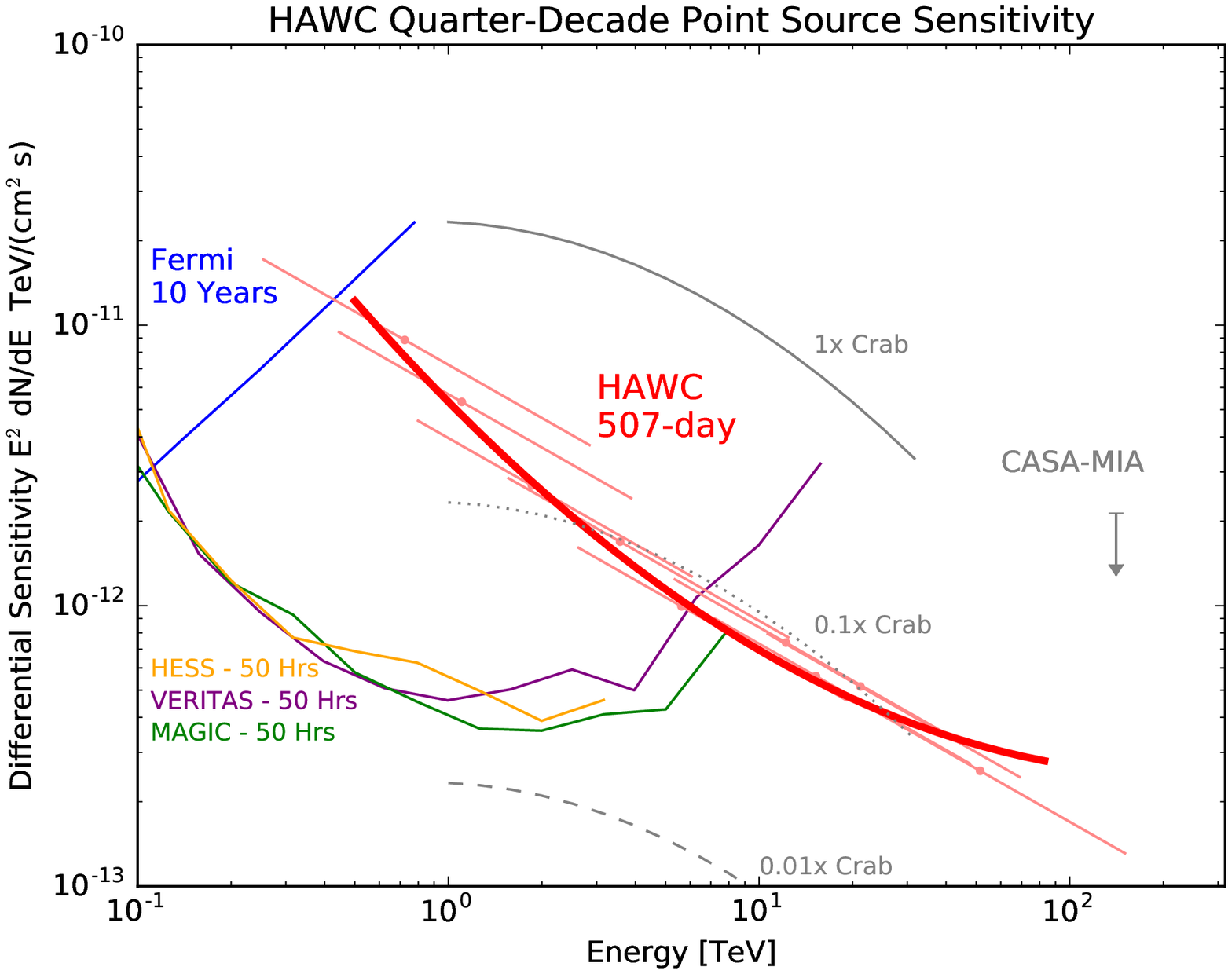}
\caption{The quasi-differential sensitivity of 
HAWC as a function
of photon energy, compared to existing IACTS \citep{veritasdifferential,hesscrab2015,magicdifferential}
and Large Area Telescope on the Fermi Gamma-Ray Space Telescope  
\footnote{Pass 8 Sensitivity: https://www.slac.stanford.edu/exp/glast/groups/canda/lat\_Performance.htm}.
We show the flux, assuming a source with a differential energy spectrum
$E^{-2.63}$, required to produce a 5$\sigma$ detection 50\% of the time. 
This flux is shown in light red for each of the
9 $\mathcal{B}$ bins, with a width in energy corresponding to the
central 68\% containment energies in each bin. These values are adjusted
to find the equivalent quarter-decade-separated flux sensitivities,
and a fit to these values is shown in dark red. 
The 507-day observation of HAWC corresponds to $\sim$3000 hours
of a source at a declination of 22$^\circ$ within HAWC's field-of-view.
HAWC's one-year sensitivity surpasses a 50-hour observation
by current-generation
IACTs at $\sim$ 10 TeV. }
\label{fig:diffsensi}
\end{figure}

Figure \ref{fig:diffsensi} shows the computed differential 
sensitivity of HAWC to sources at the declination of the Crab
utilizing the procedure of \citep{sensipaper} with the analysis 
presented here. A point source of differential photon 
spectrum $E^{-2.63}$ is simulated and fitted using 
the full likelihood fit. The flux required to be detected
at 5$\sigma$ 50\% of the time is shown for each bin, $\mathcal{B}$. 
The lines for each $\mathcal{B}$ are shown with a width
corresponding to the width required to contain 68\% of the
events under the $E^{-2.63}$ hypothesis. A correction is made to
adjust the $\mathcal{B}$ separation to a quarter decade in true energy,
and the result is fitted. 
The sensitivity prediction from \citep{sensipaper} suffered from 
uncertain background at the highest energies. Now, with more than
a year of data,
we know the background precisely and can 
set the cuts appropriately. This has resulted in a more accurate
(and more sensitive) analysis above 10 TeV. Below about 1 TeV, 
for a number of reasons, the sensitivity is somewhat worse than 
predicted in the original study. The background is larger than 
the original simulation-only prediction. Furthermore,
in the current analysis we employ a relatively high cut (defined
by $\mathcal{B}=1$)
so that improperly modeled noise can be neglected.

\subsection{Anticipated Improvements}
\label{sec:improvements}

The main limitation of this analysis is the reliance on the 
number of PMTs, used for the definition of $\mathcal{B}$, to 
simultaneously constrain the energy of photons, the angular resolution,
and photon/hadron efficiency. Figure \ref{fig:energy} shows
that this is a poor energy estimation with each bin $\mathcal{B}$ 
spanning roughly an order of magnitude of energy. More critically, an
overhead $\sim$10 TeV photon can trigger nearly every PMT in HAWC if
the core lands near the center of the detector. Consequently
$\mathcal{B}=9$ is an overflow bin of everything above 10 TeV. These
limitations can be removed with an event parameter that accounts for the
light level in the event and the specific geometry and inclination 
angle of events. Approaches like this are under development. 
The planned deployment of a sparse ``outrigger'' array should 
further increase the sensitivity to 
photons above 10 TeV \citep{hawcoutriggers}.

Additionally, the principal systematic error (the modeling of late light)
is conservatively estimated here and is being studied using the 
calibration system. It is likely that the effects of late light will
be better modeled in the future. 

Finally, the threshold for this analysis is established by including
only
events where more than 6.7\% of the PMTs detect light.
The typical number of live, calibrated PMTs is $\sim$1000,
corresponding to a threshold of $\sim$70 PMTs. Events with
20--30 PMTs could be reconstructed if the noise could be confidently identified.
A relatively high event size threshold is used in this analysis to reduce its 
dependence on the modeling of noise hits. Planned improvements in the 
modeling should lower the energy threshold of the spectrum analysis in 
future studies.

The HAWC instrument is performing well with survey sensitivity
exceeding current-generation instruments above 10 TeV, sensitivity
which HAWC maintains across much of its field-of-view. The all-sky
survey conducted by HAWC 
probes unique flux space and reveals the highest-energy 
photon sources in the northern sky. Understanding the Crab gives
confidence in the survey results.

\acknowledgments

We	acknowledge	the	support	from:	the	US	National	Science	Foundation	(NSF);	the	
US	Department	of	Energy	Office	of	High-Energy	Physics;	the	Laboratory	Directed	
Research	and	Development	(LDRD)	program	of	Los	Alamos	National	Laboratory;	
Consejo	Nacional	de	Ciencia	y	Tecnolog\'{\i}a	(CONACyT),	M{\'e}xico	(grants	
271051,	232656,	260378,	179588,	239762,	254964,	271737,	258865,	243290,	
132197);	L'OREAL	Fellowship	for Women	in	Science	2014;	Red	HAWC,	M{\'e}xico;	
DGAPA-UNAM	(grants	RG100414,	IN111315,	IN111716-3,	IA102715,	109916);	
VIEP-BUAP;	PIFI	2012,	2013,	PROFOCIE	2014,	2015; the	University	of	Wisconsin	
Alumni	Research	Foundation;	the	Institute	of	Geophysics,	Planetary	Physics,	and	
Signatures	at	Los	Alamos	National	Laboratory;	Polish	Science	Centre	grant	DEC-
2014/13/B/ST9/945;	Coordinaci{\'o}n	de	la	Investigaci{\'o}n	Cient\'{\i}fica	de	la	
Universidad	Michoacana.

\bibliography{bibliography}

\begin{thebibliography}{}
\expandafter\ifx\csname natexlab\endcsname\relax\def\natexlab#1{#1}\fi
\providecommand{\url}[1]{\href{#1}{#1}}

\bibitem[{{Abdo} {et~al.}(2009)}]{milagrolsa}
{Abdo}, A.~A., {et~al.} 2009, ApJ, 698, 2121

\bibitem[{{Abdo} {et~al.}(2011)}]{fermiflare}
---. 2011, Science, 331, 739

\bibitem[{{Abdo} {et~al.}(2012)}]{milagrocrabspectrum}
---. 2012, \apj, 750, 63

\bibitem[{{Abeysekara} {et~al.}(2012)}]{hawcgrbsensi}
{Abeysekara}, A.~U., {et~al.} 2012, Astroparticle Physics, 35, 641

\bibitem[{{Abeysekara} {et~al.}(2013)}]{sensipaper}
---. 2013, Astropart. Phys, 50, 26

\bibitem[{{Abeysekara} {et~al.}(2016)}]{hawc111bsl}
---. 2016, \apj, 817, 3

\bibitem[{{Abramowski} {et~al.}(2014)}]{hesscrabflarelimits}
{Abramowski}, A., {et~al.} 2014, \aap, 562, L4

\bibitem[{Agostinelli {et~al.}(2003)}]{geant4}
Agostinelli, S., {et~al.} 2003, Nucl. Inst. Meth. Phys. Res. A, 506, 250

\bibitem[{{Aharonian} {et~al.}(2004)}]{hegracrab}
{Aharonian}, F., {et~al.} 2004, \apj, 614, 897

\bibitem[{{Aharonian} {et~al.}(2006)}]{hesscrab}
---. 2006, \aap, 457, 899

\bibitem[{{Aielli} {et~al.}({2010})}]{argoflare}
{Aielli}, G., {et~al.} {2010}, {ATel}, {2921}

\bibitem[{{Aleksi{\'c}} {et~al.}(2015)}]{magiccrabnebula}
{Aleksi{\'c}}, J., {et~al.} 2015, Journal of High Energy Astrophysics, 5, 30

\bibitem[{{Aleksi{\'c}} {et~al.}(2016){Aleksi{\'c}}, {Ansoldi}, {Antonelli},
  {Antoranz}, {Babic}, {Bangale}, {Barcel{\'o}}, {Barrio}, {Becerra
  Gonz{\'a}lez}, {Bednarek}, {Bernardini}, {Biasuzzi}, {Biland}, {Bitossi},
  {Blanch}, {Bonnefoy}, {Bonnoli}, {Borracci}, {Bretz}, {Carmona}, {Carosi},
  {Cecchi}, {Colin}, {Colombo}, {Contreras}, {Corti}, {Cortina}, {Covino}, {Da
  Vela}, {Dazzi}, {De Angelis}, {De Caneva}, {De Lotto}, {de O{\~n}a Wilhelmi},
  {Delgado Mendez}, {Dettlaff}, {Dominis Prester}, {Dorner}, {Doro}, {Einecke},
  {Eisenacher}, {Elsaesser}, {Fidalgo}, {Fink}, {Fonseca}, {Font}, {Frantzen},
  {Fruck}, {Galindo}, {Garc{\'{\i}}a L{\'o}pez}, {Garczarczyk}, {Garrido
  Terrats}, {Gaug}, {Giavitto}, {Godinovi{\'c}}, {Gonz{\'a}lez Mu{\~n}oz},
  {Gozzini}, {Haberer}, {Hadasch}, {Hanabata}, {Hayashida}, {Herrera},
  {Hildebrand}, {Hose}, {Hrupec}, {Idec}, {Illa}, {Kadenius}, {Kellermann},
  {Knoetig}, {Kodani}, {Konno}, {Krause}, {Kubo}, {Kushida}, {La Barbera},
  {Lelas}, {Lemus}, {Lewandowska}, {Lindfors}, {Lombardi}, {Longo},
  {L{\'o}pez}, {L{\'o}pez-Coto}, {L{\'o}pez-Oramas}, {Lorca}, {Lorenz},
  {Lozano}, {Makariev}, {Mallot}, {Maneva}, {Mankuzhiyil}, {Mannheim},
  {Maraschi}, {Marcote}, {Mariotti}, {Mart{\'{\i}}nez}, {Mazin}, {Menzel},
  {Miranda}, {Mirzoyan}, {Moralejo}, {Munar-Adrover}, {Nakajima}, {Negrello},
  {Neustroev}, {Niedzwiecki}, {Nilsson}, {Nishijima}, {Noda}, {Orito},
  {Overkemping}, {Paiano}, {Palatiello}, {Paneque}, {Paoletti}, {Paredes},
  {Paredes-Fortuny}, {Persic}, {Poutanen}, {Prada Moroni}, {Prandini},
  {Puljak}, {Reinthal}, {Rhode}, {Rib{\'o}}, {Rico}, {Rodriguez Garcia},
  {R{\"u}gamer}, {Saito}, {Saito}, {Satalecka}, {Scalzotto}, {Scapin},
  {Schultz}, {Schlammer}, {Schmidl}, {Schweizer}, {Shore}, {Sillanp{\"a}{\"a}},
  {Sitarek}, {Snidaric}, {Sobczynska}, {Spanier}, {Stamerra}, {Steinbring},
  {Storz}, {Strzys}, {Takalo}, {Takami}, {Tavecchio}, {Tejedor}, {Temnikov},
  {Terzi{\'c}}, {Tescaro}, {Teshima}, {Thaele}, {Tibolla}, {Torres}, {Toyama},
  {Treves}, {Vogler}, {Wetteskind}, {Will}, \& {Zanin}}]{magicdifferential}
{Aleksi{\'c}}, J., {Ansoldi}, S., {Antonelli}, L.~A., {et~al.} 2016,
  Astroparticle Physics, 72, 76

\bibitem[{{Aliu} {et~al.}(2014)}]{veritascrabflarelimits}
{Aliu}, E., {et~al.} 2014, \apjl, 781, L11

\bibitem[{{Amenomori} {et~al.}(2009)}]{tibetcrab}
{Amenomori}, M., {et~al.} 2009, \apj, 692, 61

\bibitem[{{Amenomori} {et~al.}(2015)}]{tibet100tevul}
---. 2015, \apj, 813, 98

\bibitem[{{Atkins} {et~al.}(2003)}]{milagrocrab}
{Atkins}, R., {et~al.} 2003, \apj, 595, 803

\bibitem[{{Atoyan} \& {Aharonian}(1996)}]{crabinversecompton}
{Atoyan}, A.~M., \& {Aharonian}, F.~A. 1996, \mnras, 278, 525

\bibitem[{{Bartoli} {et~al.}(2015)}]{argocrab}
{Bartoli}, B., {et~al.} 2015, \apj, 798, 119

\bibitem[{{Celik}(2008)}]{veritascrab}
{Celik}, O. 2008, in International Cosmic Ray Conference, Vol.~2, International
  Cosmic Ray Conference, 847--850

\bibitem[{{Comella} {et~al.}(1969){Comella}, {Craft}, {Lovelace}, \&
  {Sutton}}]{oldcrabpulsarradio}
{Comella}, J.~M., {Craft}, H.~D., {Lovelace}, R.~V.~E., \& {Sutton}, J.~M.
  1969, \nat, 221, 453

\bibitem[{{Greisen}(1960)}]{nkgfunction}
{Greisen}, K. 1960, Annu. Rev. of Nucl. Part. Sci., 10, 63

\bibitem[{{Heck} {et~al.}(1998){Heck}, {Knapp}, {Capdevielle}, {Schatz}, \&
  {Thouw}}]{corsika}
{Heck}, D., {Knapp}, J., {Capdevielle}, J.~N., {Schatz}, G., \& {Thouw}, T.
  1998, {CORSIKA: a Monte Carlo code to simulate extensive air showers.}

\bibitem[{Holler {et~al.}(2016)}]{hesscrab2015}
Holler, M., {et~al.} 2016, PoS, ICRC2015, 847

\bibitem[{Lauer(2013)}]{calibrationicrc2013}
Lauer, R. 2013, in {Proceedings, 33rd International Cosmic Ray Conference
  (ICRC2013): Rio de Janeiro, Brazil, July 2-9, 2013}, 0566

\bibitem[{{Mart{\'{\i}}n} {et~al.}(2012){Mart{\'{\i}}n}, {Torres}, \&
  {Rea}}]{crabmodeling}
{Mart{\'{\i}}n}, J., {Torres}, D.~F., \& {Rea}, N. 2012, \mnras, 427, 415

\bibitem[{Meagher(2016)}]{veritascrab2015}
Meagher, K. 2016, PoS, ICRC2015, 792

\bibitem[{Park(2016)}]{veritasdifferential}
Park, N. 2016, PoS, ICRC2015, 771

\bibitem[{Sandoval(2016)}]{hawcoutriggers}
Sandoval, A. 2016, PoS, ICRC2015, 977

\bibitem[{Solares {et~al.}(2016)Solares, Gerhardt, Hui, Lauer, Ren,
  Salesa~Greus, \& Zhou}]{calibrationicrc2015}
Solares, H.~A., Gerhardt, M., Hui, C.~M., {et~al.} 2016, PoS, ICRC2015, 997

\bibitem[{{Tanimori} {et~al.}(1998)}]{cangaroocrab}
{Tanimori}, T., {et~al.} 1998, \apjl, 492, L33

\bibitem[{{Tavani} {et~al.}(2011)}]{agilecrabflare}
{Tavani}, M., {et~al.} 2011, Science, 331, 736

\bibitem[{{Weekes} {et~al.}(1989)}]{whipplecrabdiscovery}
{Weekes}, T.~C., {et~al.} 1989, \apj, 342, 379

\bibitem[{{Wilks}(1938)}]{wilkstheorem}
{Wilks}, S.~S. 1938, {Ann. Math. Statist.}, 9, 60

\bibitem[{Younk {et~al.}(2016)Younk, Lauer, Vianello, Harding, Ayala~Solares,
  \& Zhou}]{liff}
Younk, P.~W., Lauer, R.~J., Vianello, G., {et~al.} 2016, PoS, ICRC2015, 948

\end{thebibliography}

\end{document}